\documentclass[11pt,a4paper]{article}
\pdfoutput=1

\usepackage{jheppub}
\usepackage{amsmath, latexsym, amssymb, hyperref, graphicx, color, multirow, makecell, diagbox}
\usepackage{tabularx}
\usepackage[T1]{fontenc}

\usepackage{dcolumn}
\usepackage{xspace}
\usepackage{comment}

\usepackage{listings}
\lstdefinestyle{mystyle}{mathescape,basicstyle=\small\ttfamily,frame=leftline,aboveskip=4mm,belowskip=4mm,xleftmargin=20pt,framexleftmargin=10pt,numbers=none,framerule=2pt,abovecaptionskip=0.0mm,belowcaptionskip=3.5mm}
\usepackage{float}
\usepackage{epsfig}
\usepackage{longtable}
\usepackage{rotating}
\usepackage{ulem}
\usepackage{txfonts}
\usepackage{upgreek}
\usepackage{xcolor}
\usepackage{mathtools}
\usepackage{tabularx}
\usepackage{booktabs}
\usepackage{appendix}
\usepackage{pifont}% http://ctan.org/pkg/pifont
\usepackage{comment}

\usepackage{soul} % Striking text
\usepackage[capitalize]{cleveref}
\definecolor{nicered}{rgb}{.7,.1,.1}
\definecolor{nicegreen}{rgb}{.1,.5,.1}
\definecolor{darkblue}{rgb}{0,0,.5}
\hypersetup{colorlinks, citecolor=nicegreen,linkcolor=nicered, urlcolor=darkblue}
\usepackage{multirow}
\usepackage{slashed}
\usepackage{verbatim}

\usepackage{tikz}
\usetikzlibrary{positioning,arrows}
\usetikzlibrary{decorations.pathmorphing}
\usetikzlibrary{decorations.markings}
\usetikzlibrary{shapes.geometric}
\tikzset{
    vector/.style={decorate, decoration={snake}, draw},
    provector/.style={decorate, decoration={snake,amplitude=2.5pt}, draw},
    antivector/.style={decorate, decoration={snake,amplitude=-2.5pt}, draw},
    fermion/.style={draw=black, line width=0.4mm,
      postaction={decorate},decoration={markings,mark=at position .55
        with {\arrow[draw=black]{>}}}},
    fermionbar/.style={draw=black, postaction={decorate},
                       decoration={markings,mark=at position .55 with {\arrow[draw=black]{<}}}},
    fermionnoarrow/.style={draw=black},
    top/.style={draw=black, line width=1mm,
      postaction={decorate},decoration={markings,mark=at position .55
        with {\arrow[draw=black]{>}}}},
    gluon/.style={decorate, draw=black,decoration={coil,amplitude=4pt, segment length=5pt}},
    photon/.style={decorate, draw=red,decoration={snake,amplitude=5pt, segment length=9pt}},
    scalar/.style={dashed,draw=black,
      postaction={decorate},decoration={markings,mark=at position .55
        with {\arrow[draw=black]{>}}}},
    scalarbar/.style={dashed,draw=black,
      postaction={decorate},decoration={markings,mark=at position .55
        with {\arrow[draw=black]{<}}}},
    scalarnoarrow/.style={dashed,draw=black},
    electron/.style={draw=black,
      postaction={decorate},decoration={markings,mark=at position .55
        with {\arrow[draw=black]{>}}}},
    bigvector/.style={decorate, decoration={snake,amplitude=4pt}, draw},
}

\def\ttt#1{\texttt{\small #1}}

\newcommand{\red}[1]{\textcolor[rgb]{1,0,0}{#1}}

\newcommand{\forestgreen}[1]{\textcolor[rgb]{0.25,0.53,0.22}{#1}}

\newcommand{\cmark}{\text{\forestgreen{\ding{51}}}}
\newcommand{\xmark}{\text{\red{\ding{55}}}}
\newcommand{\halfcmark}{\textcolor{orange}{\ding{51}}{\small\textcolor{orange}{\kern-0.7em\ding{55}}}}

\newcommand{\pp}{p-p}

\providecommand{\pA}{p-A}
\providecommand{\pPb}{p-Pb}
\providecommand{\AaAa}{A-A}
\providecommand{\PbPb}{Pb-Pb}

\newcommand{\epem}{e^+e^-}

\newcommand{\mumu}{\mu^+\mu^-}
\newcommand{\fourmuon}{\mu^+\mu^+\mu^-\mu^-}
\newcommand{\llvv}{\mu^+\nu_\mu e^-\bar{\nu}_e}
\newcommand{\tautau}{\tau^+\tau^-}
\newcommand{\WpWm}{W^+W^-}
\newcommand{\ttbar}{t\bar{t}}

\newcommand{\gaga}{\gamma\gamma}

\newcommand{\nmax}{{\tt n}_{\tt max}}
\newcommand{\mmax}{{\tt m}_{\tt max}}
\def\beq{\begin{equation}}
\def\eeq{\end{equation}}
\def\beqn{\begin{eqnarray}}
\def\eeqn{\end{eqnarray}}

\newcommand\ident{{\cal I}}

\newcommand{\sqrts}{\sqrt{s}}

\newcommand{\sqrtsnn}{\sqrt{s_{_\text{NN}}}}

\newcommand{\MSbar}{{\rm \overline{MS}}}

\newcommand{\ABgagaX}{A$_1$ \,A$_2$\,$\xrightarrow{\gaga}$ A$_1$\, $X$ \,A$_2$}
\providecommand{\sigmagagaX}{\sigma_{\gaga\to X}}

\newcommand{\helaconia}{\textsc{HELAC-Onia}}
\newcommand{\madgraph}{\textsc{MadGraph5\_aMC@NLO}}
\newcommand{\mgshort}{\textsc{MG5\_aMC}}
\newcommand{\gammaUPC}{\ttt{gamma-UPC}}
\newcommand{\upcgen}{\ttt{UPCgen}}
\newcommand{\cepgen}{\ttt{cepgen}}
\newcommand{\starlight}{\textsc{Starlight}}
\newcommand{\superchic}{\textsc{Superchic}}
\newcommand{\fastjet}{\textsc{FastJet}}
\newcommand{\mcatnlo}{\textsc{MC@NLO}}
\newcommand{\madfks}{\textsc{MadFKS}}
\newcommand{\madloop}{\textsc{MadLoop}}
\newcommand{\cuttools}{\textsc{CutTools}}
\newcommand{\samurai}{\textsc{Samurai}}
\newcommand{\ninja}{\textsc{Ninja}}
\newcommand{\pjfry}{\textsc{PJFry++}}
\newcommand{\iregi}{\textsc{IREGI}}
\newcommand{\golem}{\textsc{Golem95}}
\newcommand{\collier}{\textsc{Collier}}
\newcommand{\openloops}{\textsc{OpenLoops}}
\newcommand\prompt{{\tt MG5\_aMC>}}
\newcommand{\taggeda}{{\tt !a!}}
\newcommand{\pythia}{\textsc{Pythia}}

\def\gev{\ \mathrm{GeV}}

\usepackage{xspace}
\newcommand*{\eg}{e.g.,\@\xspace}
\newcommand*{\ie}{i.e.,\@\xspace}
\newcommand*{\ala}{\`a la\@\xspace}
\newcommand*{\cm}{c.m.\@\xspace}

\begin{document}

\title{Automated next-to-leading order QCD and electroweak predictions of photon-photon processes in ultraperipheral collisions}

\author{Hua-Sheng Shao and Lukas Simon}
\emailAdd{huasheng.shao@lpthe.jussieu.fr,lsimon@lpthe.jussieu.fr}
\affiliation{Laboratoire de Physique Th\'eorique et Hautes Energies (LPTHE), UMR 7589, Sorbonne Universit\'e et CNRS, 4 place Jussieu, 75252 Paris Cedex 05, France
}

\preprint{}

% ********** Abstract **********

\abstract{We present automated next-to-leading order QCD and/or electroweak (EW) predictions for photon-photon processes in ultraperipheral high-energy collisions of protons and ions, extending the capabilities of the \madgraph\ framework together in combination with the \gammaUPC\ code. Key aspects of this extension are discussed. We compute QCD and/or EW quantum corrections for several phenomenologically interesting processes at LHC and FCC-hh energies.}

\maketitle

\section{Introduction}

The electromagnetic field of a relativistic charged particle can be interpreted as a flux of quasi-real photons~\cite{Brodsky:1971ud,Budnev:1975poe}, with photon energies $E_\gamma$ and number densities $N_\gamma$ scaling proportionally to the Lorentz factor $\gamma_{\mathrm{L}}$ and the squared charge of the beam particle, $Z^2$, respectively. Although the study of photon-photon processes at high-energy accelerators dates back to more than thirty years~\cite{Morgan:1994ip,Whalley:2001mk,Przybycien:2008zz}, significant experimental and theoretical advances have been made in the past two decades, particularly in collisions involving proton and ion beams. This progress has been driven by the high energies and luminosities accessible at the BNL Relativistic Heavy-Ion Collider (RHIC)~\cite{Bertulani:2005ru} and the CERN Large Hadron Collider (LHC)~\cite{Baltz:2007kq,deFavereaudeJeneret:2009db}. Studies of $\gamma\gamma$ processes in proton-proton (\pp), proton-nucleus (\pA), and nucleus-nucleus (\AaAa) ultraperipheral collisions (UPCs) have greatly enriched the LHC physics program, both in searches for beyond-the-Standard Model (BSM) signatures and in precision tests of the Standard Model (SM)~\cite{Shao:2022cly}. These include searches for axion- or graviton-like particles in light-by-light scattering ($\gaga\to\gaga$)~\cite{Knapen:2016moh,ATLAS:2017fur,CMS:2018erd,ATLAS:2019azn,ATLAS:2020hii,CMS:2024bnt,Harland-Lang:2022jwn,dEnterria:2023npy}, searches for magnetic monopole pair production via $\gaga$ fusion reactions~\cite{MoEDAL:2021vix,MoEDAL:2024wbc,ATLAS:2024nzp}, constraints on anomalous quartic gauge couplings from $\gaga\to\WpWm$~\cite{CMS:2013hdf,CMS:2016rtz,ATLAS:2016lse,ATLAS:2020iwi,CMS:2022dmc}, and measurements of the anomalous magnetic moment of the tau lepton, $a_\tau=(g-2)_\tau/2$, through $\gaga\to \tautau$ reactions~\cite{ATLAS:2022ryk,CMS:2022arf,CMS:2024qjo,Shao:2023bga}.

While two-photon induced channels generally contribute to inclusive production processes at hadron colliders, they are usually strongly suppressed compared to other partonic initial states, since the (inelastic) photon density is only about $1\%$ of the quark densities~\cite{Manohar:2016nzj,Manohar:2017eqh}. In heavy-ion collisions, coherent photon collisions are typically more likely than incoherent ones, since the number density of the former scales as $Z^2$, whereas the incoherent sum of charged particles in a nucleus contributes only as $Z$. Consequently, most studies of $\gaga$ processes have been conducted in UPCs, where the transverse separation between the two hadron beams exceeds twice their transverse radii. In this regime, the beam particles, whether protons or ions, coherently emit quasi-real photons with virtualities negligible compared to the energy scale of the reactions of interest, fully justifying the use of the equivalent photon approximation (EPA)~\cite{Fermi:1924tc,vonWeizsacker:1934nji,Williams:1934ad}. The requirement of coherent photon ($\gamma_{\mathrm{coh}}$) emission from the hadron charge distribution imposes very low maximum photon virtualities, constrained by $Q_{\gamma_{\mathrm{coh}}}<R_{\mathrm{A}}^{-1}$, where $R_{\mathrm{A}}$ is the hadron radius. This translates to $Q_{\gamma_{\mathrm{coh}}}<280$ MeV for protons ($R_{\mathrm{p}}\approx 0.7$ fm) and $Q_{\gamma_{\mathrm{coh}}}<28$ MeV for lead nuclei ($R_{\mathrm{Pb}}\approx 7$ fm). Since the hadrons remain intact--except for possible ion excitations leading to forward neutron emission~\cite{Crepet:2024dvv}--UPCs produce exceptionally clean event topologies, characterized by final-state particles detected centrally with (almost) empty forward regions. In the following discussions, we refer to the initial-state photons $\gamma$ exclusively as coherent photons ($\gamma_{\mathrm{coh}}$) and omit the subscript ``$\mathrm{coh}$" for brevity.

Experimental measurements of several $\gaga$ processes with relatively large cross sections, such as the Breit–Wheeler~\cite{Breit:1934zz} process $\gaga\to \epem$~\cite{Vane:1992ms,CERESNA45:1994cpb,STAR:2004bzo,CDF:2006apx,PHENIX:2009xtn,CMS:2012cve,ALICE:2013wjo,ATLAS:2015wnx,CMS:2018uvs,STAR:2019wlg,ATLAS:2020mve,CMS:2024bnt}, dimuon production $\gaga\to \mumu$~\cite{CMS:2011vma,ATLAS:2015wnx,ATLAS:2017sfe,CMS:2018uvs,ATLAS:2020epq,ATLAS:2020mve,CMS:2020skx}, tau-lepton pair production $\gaga\to \tautau$~\cite{ATLAS:2022ryk,CMS:2022arf,CMS:2024qjo}, and light-by-light scattering $\gaga\to\gaga$~\cite{ATLAS:2017fur,CMS:2018erd,ATLAS:2019azn,ATLAS:2020hii,CMS:2024bnt,dEnterria:2023npy}, have already entered the precision regime (with relative experimental uncertainties below $20\%$) at the LHC using Run 2 data. Furthermore, significantly more proton-proton and heavy-ion data will be collected at the LHC in the coming years, allowing for a substantial reduction in experimental uncertainties (\eg\ $\gaga\to \WpWm$) and opening new discovery channels (\eg\ $\gaga\to \ttbar$). Given these advancements, it is increasingly important to improve theoretical calculations and event simulations for two-photon UPC processes by incorporating higher-order QCD and electroweak (EW) corrections in matrix elements and/or by refining the modeling of effective photon-photon luminosities. Regarding photon-photon luminosities, several phenomenological implementations are available in public tools, such as \starlight~\cite{Klein:2016yzr}, \superchic~\cite{Harland-Lang:2020veo}, \ttt{FPMC}~\cite{Boonekamp:2011ky}, \cepgen~\cite{Forthomme:2018ecc}, \upcgen~\cite{Burmasov:2021phy}, and \gammaUPC~\cite{Shao:2022cly}. Ideally, the effective photon-photon luminosities can be significantly improved through experimental measurements, analogous to the determinations of parton distribution functions (PDFs). On the other hand, as in inclusive reactions, matrix elements can be systematically improved by incorporating higher-order quantum corrections using well-established techniques. For instance, next-to-leading order (NLO) QCD corrections to top quark pair production $\gaga\to t\bar{t}$~\cite{Shao:2022cly}, NLO QED corrections to $\gaga\to \mumu$ and $\gaga\to \tautau$~\cite{Shao:2024dmk}, NLO EW corrections to $\gaga\to \tautau$~\cite{Jiang:2024dhf}, and NLO QCD and QED corrections to $\gaga\to\gaga$~\cite{AH:2023kor,AH:2023ewe} are already known in the literature. Notably, NLO QED corrections are essential for achieving precise theoretical predictions in dimuon production and for accurately interpreting experimental measurements~\cite{Shao:2024dmk}. However, these corrections have been implemented in a process-specific manner. Consequently, it is now both timely and compelling to develop a fully general framework for NLO QCD and EW calculations of arbitrary $\gaga$ UPC processes, making automation a natural solution--analogous to what has been achieved for inclusive reaction processes within the \madgraph\ (\mgshort\ hereafter) framework~\cite{Alwall:2014hca,Frederix:2018nkq}. The well-established techniques for NLO calculations ensure the feasibility of automating $\gaga$ processes that are not loop-induced, involve only elementary particles, and do not feature excessively high final-state multiplicities. This is the primary objective of our work. Meanwhile, leading-order (LO) event simulations for any $\gaga$ processes involving elementary particles and/or quarkonia have already been automated in ref.~\cite{Shao:2022cly} by integrating the \gammaUPC\ code into \mgshort~\cite{Alwall:2014hca} and the \helaconia~\cite{Shao:2012iz,Shao:2015vga} event generators.

The rest of the paper is organized as follows. Section \ref{sec:theory} reviews the theoretical framework of two-photon processes in UPCs, covering the factorization formula, modeling of effective photon-photon luminosities, the hybrid $\alpha$ renormalization scheme, FKS subtraction of infrared (IR) divergences, mixed-coupling expansions, and the \mgshort\ framework. Section \ref{sec:syntax} describes the procedure for generating processes at NLO. In section \ref{sec:results}, we present numerical results for several processes at the LHC and the Future Circular Hadron Collider (FCC-hh)~\cite{FCC:2018vvp}. Finally, we summarize our findings in section \ref{sec:conclusion}.

\section{Theoretical framework\label{sec:theory}}

%\subsection{Factorization formula and effective photon-photon luminosities}

As mentioned earlier, the calculations in this work have been performed within the \mgshort\ framework~\cite{Alwall:2014hca}, where automated NLO EW corrections and, more generally, complete NLO predictions have been discussed in ref.~\cite{Frederix:2018nkq}. Our goal is to extend this framework to $\gaga$ processes in proton and/or ion UPCs. Under collinear factorization and using the EPA, the cross section for producing a final state $X$ via photon-photon fusion in an UPC of hadrons A$_1$ and A$_2$, with charges $Z_{1,2}$, denoted as \ABgagaX, factorizes as
% the product of the elementary cross section at a given $\gaga$ center-of-mass (\cm) energy $W_{\gaga}$, $\sigmagagaX(W_{\gaga})$, convolved with the two-photon differential distribution of the colliding beams,
\begin{equation}
\sigma(\mathrm{A}_1\; \mathrm{A}_2\,\xrightarrow{\gaga} \mathrm{A}_1 \; X \; \mathrm{A}_2)=
\int \frac{\mathrm{d}E_{\gamma_1}}{E_{\gamma_1}} \frac{\mathrm{d}E_{\gamma_2}}{E_{\gamma_2}} \, \frac{\mathrm{d}^2N^{(\mathrm{A}_1\mathrm{A}_2)}_{\gamma_1/\mathrm{Z}_1,\gamma_2/\mathrm{Z}_2}}{\mathrm{d}E_{\gamma_1}\mathrm{d}E_{\gamma_2}} \sigmagagaX(W_{\gaga}=2\sqrt{E_{\gamma_1}E_{\gamma_2}})\,,
\label{eq:two-photon}
\end{equation}
where the effective photon-photon luminosity
\begin{equation}
\frac{\mathrm{d}^2N^{(\mathrm{A}_1\mathrm{A}_2)}_{\gamma_1/\mathrm{Z}_1,\gamma_2/\mathrm{Z}_2}}{\mathrm{d}E_{\gamma_1}\mathrm{d}E_{\gamma_2}} =  \int{\mathrm{d}^2\pmb{b}_1\mathrm{d}^2\pmb{b}_2\,P_\text{no\,inel}(\pmb{b}_1,\pmb{b}_2)\,N_{\gamma_1/\mathrm{Z}_1}(E_{\gamma_1},\pmb{b}_1)N_{\gamma_2/\mathrm{Z}_2}(E_{\gamma_2},\pmb{b}_2)}\label{eq:2photonintegral}
\end{equation}
is the convolution of the two photon number densities, $N_{\gamma_1/\mathrm{Z}_1}(E_{\gamma_1},\pmb{b}_1)$ and $N_{\gamma_2/\mathrm{Z}_2}(E_{\gamma_2},\pmb{b}_2)$, at impact parameters $\pmb{b}_{1,2}$ with respect to the centers of hadrons A$_1$ and A$_2$. The factor $P_\text{no\,inel}(\pmb{b}_1,\pmb{b}_2)$ accounts for the probability of no inelastic hadronic interactions between A$_1$ and A$_2$. The \gammaUPC\ code~\cite{Shao:2022cly} provides two types of elastic/coherent photon fluxes as functions of the impact parameter: the electric-dipole (EDFF) and charge (ChFF) form factors for protons and nuclei. Generally, cross sections computed using the ChFF flux align closely with \superchic, while those based on the EDFF photon flux resemble \starlight. However, the ChFF flux is preferred for at least two reasons. First, the EDFF photon number density diverges at vanishing impact parameters, necessitating an arbitrary cutoff (typically $b=|\pmb{b}|\gtrsim R_{\mathrm{A}}$) to regulate the integral in eq.~\eqref{eq:2photonintegral}. In contrast, the ChFF photon flux remains well-behaved for all values of $b$, making it more realistic. In fact, cross sections computed with the ChFF photon flux agree better with experimental results for dilepton production processes, such as $\gaga\to \epem$~\cite{CMS:2024bnt} and $\gaga\to \mumu$~\cite{Shao:2024dmk}. For these reasons, we will use the ChFF flux throughout this paper. Finally, due to the presence of $P_\text{no\,inel}(\pmb{b}_1,\pmb{b}_2)$, the effective two-photon luminosity in eq.~\eqref{eq:2photonintegral} cannot be factorized into a simple product of two PDF-like functions.

In NLO EW calculations, a key observation for UPC $\gaga$ processes is that the initial photons should be treated as long-distance photons, rather than short-distance photons, due to their very small virtualities. This leads to at least two important consequences. First, the QED fine structure constant $\alpha$ related to vertices involving long-distance photons (either initial or final tagged~\cite{Pagani:2021iwa}) should take $\alpha(0)\approx 1/137$, while $\alpha$ for short-distance photons, as well as for other particles, should be defined in an $\MSbar$-like ultraviolet (UV) renormalization scheme, such as $\alpha(M_Z)$ or the $G_\mu$ scheme (\eg\ see section 3.3 of ref.~\cite{Shao:2025fwd}). This leads to the necessity of a hybrid $\alpha$ renormalization scheme in virtual corrections at NLO, as done in ref.~\cite{Pagani:2021iwa} for isolated photons in final states. For processes involving $n_\gamma\geq 0$ isolated photons in the final state $X$, the elementary cross section $\sigmagagaX$ must be proportional to $\alpha(0)^{n_\gamma+2}$, where the power of $2$ accounts for the two initial photons. Any additional $\alpha$ dependence in $\sigmagagaX$ should either be $\alpha_{G_\mu}\approx 1/132$ (evaluated in the $G_\mu$ scheme) or $\alpha(M_Z)\approx1/129$. 

The second aspect we need to address is related to real corrections. As in ref.~\cite{Pagani:2021iwa}, we must ensure that there are two initial photons, thereby preventing any potential initial photon splitting, $f\to \gamma f$, where $f$ is a massless charged fermion. This requires vetoing real-correction processes where the incoming particle is anything other than a photon. As a result, the IR divergence subtraction counterterms \ala FKS~\cite{Frixione:1995ms,Frixione:1997np,Frederix:2009yq} must be adjusted accordingly. This leads to setting $n_I=3$ in the local counterterms, and 
\begin{equation}
C_{\mathrm{EW}}(\gamma)=\gamma_{\mathrm{EW}}(\gamma)=\gamma^\prime_{\mathrm{EW}}(\gamma)=0
\end{equation}
for initial photons $\gamma$ in the integrated counterterms, using the notations introduced in section 3.2 of ref.~\cite{Shao:2025fwd}. Similar adjustments should be made for final isolated photons. However, for $\gaga$-initiated processes with jets in the final state, we must address the initial collinear divergence arising from the splitting $\gamma \to f\bar{f}$, where $f$ is a massless charged fermion from the parton content of jets. In such cases, we must generate the underlying Born processes with $f$ or $\bar{f}$ in the initial state, despite their fluxes being zero, by setting $n_I=1$. Additionally, we introduce the EW PDF counterterm by defining
\begin{equation}
K_{\ident\gamma}^{(\mathrm{QCD})}(z)=0,\quad K_{\ident\gamma}^{(\mathrm{EW})}(z)=-\log{\left(\mu^2\xi_{\mathrm{A}}^2R_{\mathrm{A}}^2\right)}P_{\ident\gamma}^{(\mathrm{EW})}(z,0)\,,\label{eq:Kdef4photon}
\end{equation}
as given in eq.~(3.2.22) of ref.~\cite{Shao:2025fwd}, where $\mu$ is the dimensional regularization scale, $\xi_{\mathrm{A}}$ is an arbitrary $\mathcal{O}(1)$ factor, $R_{\mathrm{A}}$ is the radius of hadron A, and $P_{\ident \gamma}^{\mathrm{(EW)}}(z,0)$ is the EW Altarelli-Parisi splitting kernel~\cite{Altarelli:1977zs} for $\gamma\to \ident$ in four-dimensional spacetime. The product $\xi_{\mathrm{A}}R_{\mathrm{A}}$ represents the inverse of the typical virtuality of the emitted coherent photon. We can vary $\xi_{\mathrm{A}}$ to estimate an intrinsic theory uncertainty. The presence of a nonzero $K_{\ident\gamma}^{(\mathrm{EW})}(z)$ arises because our photon fluxes are not defined in the $\MSbar$ scheme, and they do not evolve with the factorization scale. Note that, from the IR perspective (as also indicated in eq.~\eqref{eq:Kdef4photon}), such initial collinear divergences should be considered part of the EW or QED corrections to the $\gamma f$ (or $\gamma \bar{f}$)-induced underlying Born processes. However, from the coupling-counting perspective, they contribute to the NLO QCD, NLO EW, and subleading NLO corrections, which we will define below.

To clarify what we calculate in this paper, let us consider a generic observable $\Sigma$ (\eg\ a cross section for a given scattering process) that receives contributions from partonic processes involving both QCD and EW interactions. Following the notation used in refs.~\cite{Alwall:2014hca,Frixione:2014qaa,Frixione:2015zaa,Frederix:2016ost,Frederix:2018nkq}, the observable $\Sigma(\alpha_s,\alpha)$ generally admits the following Taylor expansion in the strong $\alpha_s$ and EW $\alpha$ couplings:
\begin{eqnarray}
\Sigma(\alpha_s,\alpha)&=&\alpha_s^{c_s(k_0)}\alpha^{c(k_0)}\sum_{p=0}^{+\infty}{\sum_{q=0}^{\Delta(k_0)+p}{\Sigma_{k_0+p,q}\alpha_s^{\Delta(k_0)+p-q}\alpha^q}}\nonumber\\
&=&\Sigma^{({\rm LO})}(\alpha_s,\alpha)+\Sigma^{({\rm NLO})}(\alpha_s,\alpha)+\ldots,\label{eq:Sigmaexp0}
\end{eqnarray}
where we identify the (N)LO contribution $\Sigma^{({\rm (N)LO})}$ with the $p=0$ ($1$) terms, \ie
\begin{eqnarray}
\Sigma^{({\rm LO})}(\alpha_s,\alpha)&=&\alpha_s^{c_s(k_0)}\alpha^{c(k_0)}\sum_{q=0}^{\Delta(k_0)}{\Sigma_{k_0,q}\alpha_s^{\Delta(k_0)-q}\alpha^q}\nonumber\\
&=&\Sigma_{\mathrm{LO}_1}+\ldots+\Sigma_{\mathrm{LO}_{\Delta(k_0)+1}}\,,\\
\Sigma^{({\rm NLO})}(\alpha_s,\alpha)&=&\alpha_s^{c_s(k_0)}\alpha^{c(k_0)}\sum_{q=0}^{\Delta(k_0)+1}{\Sigma_{k_0+1,q}\alpha_s^{\Delta(k_0)+1-q}\alpha^q}\nonumber\\
&=&\Sigma_{\mathrm{NLO}_1}+\ldots+\Sigma_{\mathrm{NLO}_{\Delta(k_0)+2}}\,.
\end{eqnarray}
The non-negative integers $k_0$, $c_s(k_0)$, $c(k_0)$, and $\Delta(k_0)$ are observable-dependent quantities, with the constraint $k_0=c_s(k_0)+c(k_0)+\Delta(k_0)$.  It
is customary to refer to $\Sigma_{\mathrm{NLO}_1}$ and $\Sigma_{\mathrm{NLO}_2}$ as the NLO QCD and NLO EW corrections, respectively, while $\Sigma_{\mathrm{NLO}_i}$ with $i\geq 3$ are called subleading NLO corrections, as they are typically numerically subdominant compared to the first two terms due to $\alpha_s\gg \alpha$. However, in real situations, such hierarchies may be violated (\eg see examples in chapter 4 of ref.~\cite{Shao:2025fwd}). Moreover, NLO$_i$ ($i\geq 2$) can in principle receive contributions from heavy-boson radiation (HBR) in real corrections. However, HBR is typically excluded from conventional NLO EW calculations, as it contributes at LO to a different, IR finite process, which can be experimentally reconstructed to a large extent. Therefore, we will exclude HBR in real corrections for our calculations. We will also use the following shorthand notations to present our results:
\begin{eqnarray}
\sigma_{\mathrm{LO}}&=&\Sigma^{({\rm LO})}(\alpha_s,\alpha)\,,\\
\sigma_{\mathrm{LO}_i}&=&\Sigma_{{\rm LO}_i},\quad \forall\ i\,,\\
\sigma_{\mathrm{NLO~QCD}}&=&\Sigma_{{\rm LO}_1}+\Sigma_{{\rm NLO}_1}\,,\\
\sigma_{\mathrm{NLO~EW}}&=&\Sigma_{{\rm LO}_1}+\Sigma_{{\rm NLO}_2},\\
\sigma_{\mathrm{NLO~QCD+EW}}&=&\Sigma_{{\rm LO}_1}+\Sigma_{{\rm NLO}_1}+\Sigma_{{\rm NLO}_2}\,,\\
\sigma_{\mathrm{NLO}}&=&\Sigma^{({\rm LO})}(\alpha_s,\alpha)+\Sigma^{({\rm NLO})}(\alpha_s,\alpha)\,.
\end{eqnarray}

Since our calculations are based on the \mgshort\ framework, it is useful to briefly comment on its main features. \mgshort\ is a comprehensive framework that automates the computations of both LO (tree-level or loop-induced~\cite{Hirschi:2015iia}) and NLO~\footnote{NLO computations are currently supported for processes that have nonzero tree-level LO contributions.} accurate (differential) cross sections, including their matching to parton shower (PS) Monte Carlo programs via the \mcatnlo\ method~\cite{Frixione:2002ik}. It provides all the necessary tools for studies in both SM and BSM phenomenology. The program employs the FKS subtraction method, which is automated in the \madfks\ module~\cite{Frederix:2009yq,Frederix:2016rdc} to handle IR singularities in real emission contributions. One-loop
amplitudes are computed using the \madloop\ module~\cite{Hirschi:2011pa,Alwall:2014hca}, which dynamically switches between two
Feynman integral evaluation techniques: tensor-integral and integrand-level reductions. It utilizes seven reduction tools: \cuttools~\cite{Ossola:2007ax}, \ninja~\cite{Peraro:2014cba,Hirschi:2016mdz}, \collier~\cite{Denner:2016kdg}, \samurai~\cite{Mastrolia:2010nb}, \golem~\cite{Binoth:2008uq}, \pjfry~\cite{Fleischer:2011zz}, and \iregi~\cite{Alwall:2014hca,Shao:2016knn}. The efficiency of loop numerator generation is enhanced through an in-house implementation of the \openloops\ idea~\cite{Cascioli:2011va}, which allows for dynamic switching between interfaces for different reduction programs. For matching NLO
QCD computations with parton showers, the \mcatnlo\ method~\cite{Frixione:2002ik} is used, with two new variants of the original formalism recently implemented to reduce negative weight events~\cite{Frederix:2020trv,Frederix:2023hom}. Event samples accurate at the NLO QCD level, spanning a wide range of jet multiplicities, can be obtained using
the FxFx merging method~\cite{Frederix:2012ps}.

\begin{table}[htpb!]
\centering
\tabcolsep=3.5mm
\vspace{0.2cm}
\begin{tabular}{|l|c|c|c|c|} \hline
 & \multicolumn{4}{c|}{\gammaUPC+\mgshort}\\
 & \multicolumn{2}{c|}{Processes without jets} & \multicolumn{2}{c|}{Processes with jets}\\
 & QCD ($i=1$) & EW ($i\geq 2$) & QCD ($i=1$) & EW ($i\geq 2$) \\
\hline
fLO &  \cmark &  \cmark & \cmark & \cmark \\\hline
LO+PS &  \cmark &  \cmark & \cmark & \cmark \\\hline
fNLO &  \cmark &  \cmark & \cmark & \cmark \\\hline
NLO+PS &  \cmark &  \xmark & \xmark & \xmark \\
\hline
\end{tabular}
\caption{A summary of the currently available (check mark \cmark) and unavailable (cross mark \xmark) types of computations in the \gammaUPC+\mgshort\ framework for photon-photon processes in UPCs. Here, `with' or `without' jets refers to the LO configuration; jets may still be produced at NLO via real radiation, even in processes labeled as `without jets'.\label{tab:allowedinMG}}
\end{table}

The \mgshort\ framework, together with the \gammaUPC\ code, allows us to perform the following types of calculations for arbitrary processes in proton and nuclear UPCs, following the nomenclature in ref.~\cite{Alwall:2014hca}:
\begin{itemize}
\item \textbf{fLO}: A parton-level LO computation at either tree-level or loop-induced, without PS and hadronization.
\item \textbf{fNLO}: A parton-level NLO computation for processes with non-vanishing tree-level LO scattering amplitudes. This involves both tree-level and one-loop matrix elements, but no PS is included. IR-safe observables are reconstructed using the external partons from the matrix elements.
\item \textbf{LO+PS}: Uses the matrix elements from fLO to generate unweighted events, interfacing them with PS programs. Physical observables are reconstructed using particles from Monte Carlo simulations. The interface between hard events and parton showers is straightforward, with no double-counting or singularities.
\item \textbf{NLO+PS}: Similar to LO+PS, but with NLO matrix elements. The matching of NLO-accurate events with parton showers is performed using the \mcatnlo\ formalism.
\end{itemize}
We summarize the currently available (check mark \cmark) and unavailable (cross mark \xmark) types of computations in \gammaUPC+\mgshort\ in table \ref{tab:allowedinMG}. The integer $i$ refers to which $\Sigma_{\mathrm{LO}_{i}}$ or $\Sigma_{\mathrm{NLO}_{i}}$ is considered. In conclusion, there are no constraints for fLO, LO+PS, or fNLO calculations, while NLO matching to PS is currently only possible for processes without jets~\footnote{To avoid any possible confusion, we refer to processes with or without jets as defined at LO. Jets may still be produced at NLO via real radiation, even in processes labeled as `without jets'.} and at NLO QCD accuracy.

\section{Generation syntax\label{sec:syntax}}

The process generation syntax for the new developments reported in this work largely follows that for inclusive \pp\ processes~\cite{Frederix:2018nkq} (also see section 3.7 in ref.~\cite{Shao:2025fwd}). We reuse the notation \taggeda\ (first introduced in ref.~\cite{Pagani:2021iwa}) for a tagged photon in the final state to represent a coherent photon in the initial state. In general, the generation syntax with NLO QCD and/or EW corrections is as follows:
\vskip 0.25truecm
\noindent
~~\prompt\ {\tt ~set~complex\_mass\_scheme~True}

\noindent
~~\prompt\ {\tt ~import~model~myNLOmodel\_w\_qcd\_qed-restrict\_card\_w\_a0}

\noindent
~~\prompt\ {\tt ~generate~\taggeda\ \taggeda\ > p$_1$ p$_2$ p$_3$ p$_4$ aS=$\nmax$ aEW=$\mmax$ [QCD QED]}

\vskip 0.25truecm
\noindent
where {\tt p$_i$} refers to (multi)particles in the particle spectrum of the NLO model\\
{\tt myNLOmodel\_w\_qcd\_qed}, which supports the hybrid $\alpha$ renormalization as discussed in section \ref{sec:theory}, using the restriction card option {\tt restrict\_card\_w\_a0}. A final tagged photon can be introduced by setting {\tt p$_i$}$=$\taggeda. The above commands instruct the code to compute the following LO and NLO contributions:
\beqn
&&{\rm LO}:\phantom{Naaa}
\alpha_s^p\alpha^q\,,\;\;\;\;\;p\le\nmax\,,\;\;\phantom{+1,}\;\,
q\le\mmax\,,\;\;\phantom{+1,}\;\,\,p+q=k_0\,,
\label{LOsynt}
\\*
&&{\rm NLO}:\phantom{aaa}\,
\alpha_s^p\alpha^q\,,\;\;\;\;\;p\le\nmax+1\,,\;\;\;
q\le\mmax+1\,,\;\;\;\,p+q=k_0+1\,.
\label{NLOsynt}
\eeqn
In the NLO expression, the largest power of $\alpha_s$ ($\alpha$) is exactly one unit larger than its LO counterpart due to the presence of the {\tt [QCD]} ({\tt [QED]}) keyword, which can be omitted. The values of $\nmax$ and $\mmax$ can be freely chosen by the user. If omitted, \mgshort\ generates the process with the smallest possible power of the QED coupling $\alpha$ at LO, following the  hierarchy $\alpha_s\gg \alpha$. The first command line, ``{\tt set complex\_mass\_scheme True}", tells the program to use the complex-mass scheme~\cite{Denner:1999gp,Denner:2005fg,Frederix:2018nkq} to handle intermediate resonances, which is set to {\tt False} by default.

For Born processes without jets in the final state, the parton component of \taggeda\ in both initial and final states is just a photon. However, for processes with jets in the final state,  this is not longer true for the initial \taggeda\ due to IR safety. In such cases, the initial \taggeda\ should include all massless charged fermions in the jet definition, even if their fluxes are zero. The Born matrix elements initiated by a massless fermion and a photon are required to construct FKS counterterms for initial collinear singularities, as discussed in section \ref{sec:theory}. Finally, the setup for $\gaga$ UPC processes in the {\tt run\_card.dat} file should follow the same procedure as in LO runs, as illustrated in appendix A.3 of ref.~\cite{Shao:2022cly}. Notably, the beam energy for heavy ions corresponds to the energy of the entire nucleus, not per nucleon. For processes with jets, in order to use eq.~\eqref{eq:Kdef4photon}, it is important to set {\tt pdfscheme = 7} in the {\tt run\_card.dat} file.

\section{Selected results\label{sec:results}}

In this section, we present our predictions for integrated and differential photon-photon fusion cross sections at LHC and FCC-hh energies for a wide variety of processes, with particle multiplicities ranging from $2\to 2$ to $2\to 4$ at LO. In all cases, our presented results are exclusively derived with ChFF photon fluxes. The goal of this section is to demonstrate the capabilities of our new developments in the \mgshort\ framework for enabling automated calculations of NLO QCD and EW corrections for $\gaga$ processes in UPCs.

\subsection{Setup of calculations\label{sec:setup}}

%\subsubsection{Input parameters}

We begin by outlining the baseline setup of our calculations. Any necessary modifications to this setup will be explicitly defined on a process-by-process basis. As mentioned previously, we work in a hybrid $\alpha$ renormalization scheme, using Fermi's constant from the muon decay:
\begin{equation}
G_\mu=1.16639\times 10^{-5}~\mathrm{GeV}^{-2}\,,
\end{equation}
and the value of the fine-structure constant:
\begin{equation}
\alpha(0)=\dfrac{1}{137.036}\,.    
\end{equation}
We employ a new Universal Feynman Output (UFO) format~\cite{Darme:2023jdn,Degrande:2011ua} model,\\ \texttt{loop\_qcd\_qed\_sm\_Gmu\_3FS}, which is compatible with massive charm quarks, bottom quarks, and tau leptons.\footnote{The UFO model \texttt{loop\_qcd\_qed\_sm\_Gmu\_4FS}, which allows NLO QCD and EW calculations with massive bottom quarks, was presented in ref.~\cite{Pagani:2020rsg}.} The on-shell (OS) values of $M_c,M_b$, and $M_\tau$ are summarized in table~\ref{tab:EWparaminput}, while the masses of $u$, $d$, $s$, $e^\pm$, and $\mu^\pm$ are set to zero. The OS masses of other particles, such as the top quark, $W^\pm$, $Z$, and Higgs bosons, are provided in table~\ref{tab:EWparaminput}. The particle widths are set to zero, except when calculating the following processes:
\begin{eqnarray}
\gaga&\to& \ttbar j\,,\nonumber\\
%\gaga &\to& \ttbar \bbbar\,,\nonumber\\
%\gamma\gamma&\to& t\bar{t}jj\,,\\
\gaga&\to& \fourmuon\,,\nonumber\\
\gaga&\to& \llvv\,.\label{eq:exceptionprocs}
\end{eqnarray}
In these three exceptional processes, due to the presence of OS intermediate particles, we use the complex-mass scheme and take the OS widths of $W^\pm$ and $Z$ bosons as~\cite{ParticleDataGroup:2024cfk}:
\begin{equation}
\Gamma_{W}=2.085~\mathrm{GeV}\,,\quad \Gamma_{Z}=2.4955~\mathrm{GeV}\,.
\end{equation}
In the complex-mass scheme, the OS masses and widths of $W^\pm$ and $Z$ bosons are converted into their pole masses and widths via:
\begin{equation}
\bar{M}_V=M_V/\sqrt{1+\Gamma_V^2/M_V^2}\,,\quad \bar{\Gamma}_V=\Gamma_V/\sqrt{1+\Gamma_V^2/M_V^2}\,,\quad V=W,Z\,.
\end{equation}
This results in the following pole masses and widths:
\begin{equation}
\begin{aligned}
\bar{M}_W=80.3422~\mathrm{GeV}\,,\quad & \bar{\Gamma}_W=2.0843~\mathrm{GeV}\,,\\
\bar{M}_Z=91.1539~\mathrm{GeV}\,,\quad & \bar{\Gamma}_Z=2.49457~\mathrm{GeV}\,.
\end{aligned}
\end{equation}
Additionally, we take:
\begin{equation}
\bar{M}_h=125.20~\mathrm{GeV}\,,\quad \bar{\Gamma}_h=0.00407~\mathrm{GeV}\,.
\end{equation}
The processes in eq.~\eqref{eq:exceptionprocs} may receive contributions from subsets of one-loop diagrams involving an $s$-channel Higgs boson propagator. Since their Born diagrams do not include Higgs propagators, the virtual matrix elements remain integrable even with a zero Higgs width. Therefore, we use a nonzero Higgs width solely to improve the numerical integration. Finally, the Cabibbo-Kobayashi-Maskawa (CKM) matrix~\cite{Cabibbo:1963yz,Kobayashi:1973fv} is taken to be the identity.

\begin{table}[b]
\begin{center}
\begin{tabular}{cl|cl}\toprule
Parameter & value & Parameter & value
\\\midrule
$\alpha(0)$ & \texttt{1/137.036} & $\alpha_s(M_Z)$ & \texttt{0.118}\\
$M_\tau$ & \texttt{1.77693}  & $M_c$ & \texttt{1.50}
\\
$M_b$ & \texttt{4.75}  & $M_t$ & \texttt{172.57}
\\
$M_W$ & \texttt{80.3692}  & $M_Z$ & \texttt{91.1880}
\\
$M_h$ & \texttt{125.20}  & &
\\\bottomrule
\end{tabular}
\caption{\label{tab:EWparaminput}
Parameters used in NLO calculations. All on-shell masses are given in units of $\gev$ (omitted for brevity). Widths are generally set to zero, except for the processes in eq.~\eqref{eq:exceptionprocs}, where nonzero widths for $W^\pm$, $Z$, and $h$ are used (see text for details).}
\end{center}
\end{table}

%\subsubsection{Renormalization scale}

The central renormalization scale is set to
\begin{equation}
\mu_R=H_T/2=\frac{1}{2}\sum_{i}{\sqrt{p_{T,i}^2+m_i^2}}\,,
\end{equation}
where the sum runs over all final-state partons before they are possibly clustered into jets or dressed leptons. Here, $p_{T,i}$ and $m_i$ denote the transverse momentum and invariant mass of the $i$th final-state parton, respectively. For $\gaga$ UPC processes, there is no factorization scale dependence.

%\subsubsection{Fiducial cuts}

To ensure that cross sections are defined in an IR-safe manner, we impose fiducial cuts on final-state particles as follows:
\begin{itemize}
\item \textbf{Photon isolation}: For processes with tagged final-state photon(s), these long-distance photon(s) must be defined using the photon isolation algorithm~\cite{Frixione:1998jh}. We adopt the following parameters: 
\begin{equation}
    p_T(\gamma_{\mathrm{iso}})>25~\mathrm{GeV},\quad |\eta(\gamma_{\mathrm{iso}})|<2.5, \quad R_{0,\gamma}=0.4, \quad \epsilon_\gamma=1, \quad n=1\,.
\end{equation}
\item \textbf{Photon recombination}: Dressed charged fermions are defined through the photon recombination procedure. Specifically, any photon failing the isolation cuts is recombined with the closest charged fermion if their separation satisfies $\Delta R_{f\gamma}\leq 0.1$. 
\item \textbf{Cuts on charged dressed leptons}: Charged dressed leptons ($\ell^\pm$) must satisfy: 
\begin{equation}
p_T(\ell^\pm)>10~\mathrm{GeV},\quad |\eta(\ell^\pm)|<2.5\,.
\end{equation}
Additionally, for any pair of oppositely charged, same-flavor leptons, we impose the cuts $m_{\ell^+\ell^-}>20$ GeV.
\item \textbf{Cuts on jets}: Jets ($j$) are reconstructed from gluons, (dressed) (anti)quarks, and uncombined photons using the anti-$k_T$ clustering algorithm~\cite{Cacciari:2008gp} in \fastjet~\cite{Cacciari:2011ma,Cacciari:2005hq} with a jet radius of $R=0.4$. The following cuts are applied:
\begin{equation}
p_T(j)>30~\mathrm{GeV},\quad |\eta(j)|<4.5\,.\label{eq:jetcuts}
\end{equation}
\end{itemize}

%\subsection{$\gamma\gamma\to e^+e^-$}

\subsection{$\gaga\to \mumu$\label{sec:dimuon}}

The dimuon production process, $\gaga\to \mumu$, serves as a standard candle for studying proton and ion coherent $\gamma$ fluxes, soft survival probabilities, and higher-order quantum corrections. Like their siblings electrons and positrons, muons can be copiously produced at the LHC due to their small mass. Experimentally, high-energy muons can be reconstructed even more precisely than electrons or positrons. For electrons or positrons, photon recombination is automatically incorporated during their experimental reconstruction from electromagnetic showers detected by calorimeters. In contrast, muons--being $200$ times more massive than electrons--are directly observed as bare leptons through their tracks in the muon chamber. To properly match theoretical calculations with experimental measurements of bare muons, the muon mass must be retained in perturbative calculations to regularize collinear singularities, as done in ref.~\cite{Shao:2024dmk}. However, this also introduces potentially large quasi-collinear logarithms from final-state photon radiation. To mitigate significant final-state radiation corrections,
muons are sometimes reconstructed as dressed leptons via photon recombination. The advantage of using dressed leptons instead of bare leptons is that (quasi-)collinear logarithms cancel out, making the cross section largely independent of the lepton mass. 

In this paper, we set the muon mass to zero, requiring muons to be dressed with nearby photons via photon recombination. We adopt the five-flavor scheme by setting $M_c=M_\tau=M_b=0$, following the baseline setup defined in section \ref{sec:setup}. Our calculations are performed using the following process generation commands:
\vskip 0.25truecm

\noindent
~~\prompt\ {\tt ~import~model~loop\_qcd\_qed\_sm\_Gmu\_3FS-a0}

\noindent
~~\prompt\ {\tt ~generate \taggeda\ \taggeda\ > mu+ mu- [QED]}

\vskip 0.25truecm
\noindent
We impose the conditions specified in the restriction card {\tt restrict\_a0.dat} within the model {\tt loop\_qcd\_qed\_sm\_Gmu\_3FS}. Using the notations defined in eq.~\eqref{eq:Sigmaexp0}, this process is characterized by $k_0=2$, $c_s(k_0)=0$, $c(k_0)=2$, and $\Delta(k_0)=0$. Consequently, the only contributions are a single LO term, $\Sigma_{\mathrm{LO}_1}$, at $\mathcal{O}(\alpha^2)$, and a single NLO term, $\Sigma_{\mathrm{NLO}_2}$, at $\mathcal{O}(\alpha^3)$. The NLO QCD contribution $\Sigma_{\mathrm{NLO}_1}$ at $\mathcal{O}(\alpha_s\alpha^2)$ vanishes.

\begin{table}[t]
\centering
\tabcolsep=3.5mm
\vspace{0.2cm}
\begin{tabular}{l|c|c|c} \hline
Process: $\gaga\to \mumu$ & \multicolumn{3}{c}{\gammaUPC+\mgshort} \\
Colliding system, \cm\ energy & $\sigma_{\mathrm{LO}}$ & $\sigma_{\mathrm{NLO}\ \mathrm{EW}}$ & $\sigma_{\mathrm{NLO}\ \mathrm{EW}}/\sigma_{\mathrm{LO}}$ \\ \hline
\pp\ at 13~TeV &  $1.0234(1)$ pb &  $0.98828(8)$ pb & 0.966 \\
\pp\ at 13.6~TeV &  $1.0473(1)$ pb &  $1.0114(1)$ pb & 0.966 \\
\pp\ at 14~TeV &  $1.0629(1)$ pb &  $1.0265(1)$ pb & 0.966 \\
\pPb\ at 8.8~TeV &  $2.0361(2)$ nb &  $1.9594(2)$ nb & 0.962 \\
\PbPb\ at 5.52~TeV &  $2.8861(2)$ $\mu$b &  $2.7614(2)$ $\mu$b & 0.957 \\
\hline
\pp\ at 100~TeV &  $2.3699(8)$ pb &  $2.2926(9)$ pb & 0.967 \\
\pPb\ at 62.8~TeV &  $7.953(3)$ nb &  $7.680(3)$ nb & 0.966 \\
\PbPb\ at 39.4~TeV &  $25.547(8)$ $\mu$b &  $24.631(8)$ $\mu$b & 0.964 \\
\hline
\end{tabular}
\caption{The integrated fiducial dimuon cross sections (within the fiducial cuts defined in section \ref{sec:setup}) at LO and NLO accuracy at LHC and FCC-hh energies, computed using the ChFF $\gamma$ fluxes available in \gammaUPC. The last column also shows the corresponding $K$ factors. Numbers in parentheses indicate statistical errors from numerical integrations, which may affect the last decimal place.\label{tab:dimuonxs}}
\end{table}

 We emphasize that our goal here is not to perform a data-theory comparison, which has already been done in ref.~\cite{Shao:2024dmk}. Instead, we aim to illustrate the typical size of NLO EW corrections for this process, which is relevant at the LHC and FCC-hh. The phase-space integrated fiducial cross sections for $\gaga\to\mumu$ at both LO and NLO EW in \pp, \pPb, and \PbPb\ UPCs at the LHC and FCC-hh are presented in table \ref{tab:dimuonxs}. In \pp\ collisions at center-of-mass (\cm) energies of 13-14 TeV, the fiducial cross sections are approximately $1$ pb. These increase significantly in heavy-ion collisions due to the much larger photon fluxes. For instance, the dimuon production rate in \pPb\ (\PbPb) collisions at a nucleon-nucleon \cm\ energy of $\sqrtsnn=8.8$ ($5.52$) TeV is roughly 2,000 ($3\cdot 10^6$) times higher than in \pp. The size of NLO EW corrections varies only mildly across different colliding systems, as shown in the last column of table \ref{tab:dimuonxs} for the $K$ factors ($K=\sigma_{\mathrm{NLO}\ \mathrm{EW}}/\sigma_{\mathrm{LO}}$). These corrections reduce the LO cross sections by approximately $3.4\%$ in \pp\ and $4.3\%$ in \PbPb. Although the absolute cross sections depend on the choice of photon fluxes (ChFF versus EDFF), the $K$ factors are nearly independent of the photon flux choice--an observation also noted in ref.~\cite{Shao:2024dmk}.

\begin{figure}
    \includegraphics[width=0.49\textwidth]{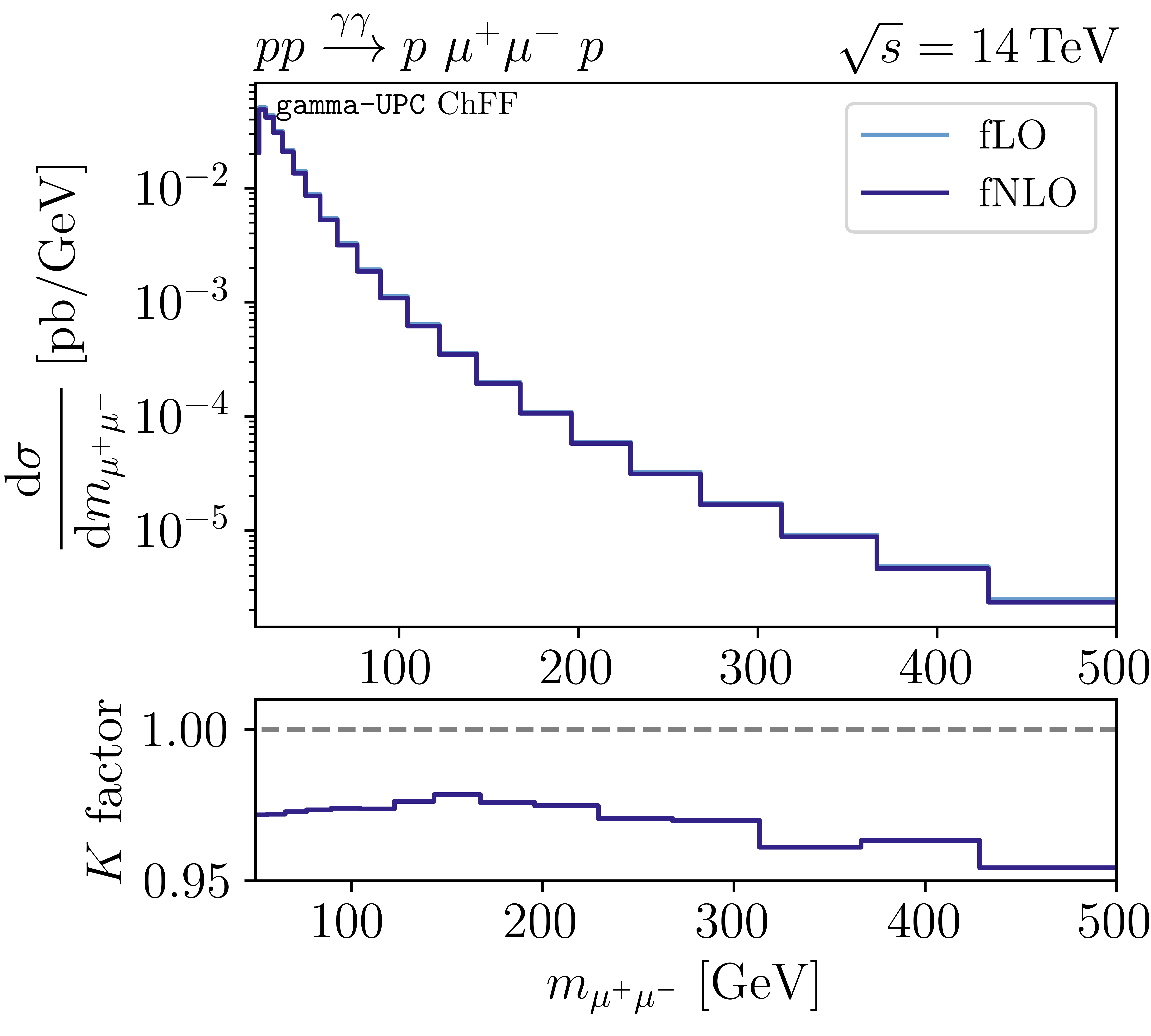}
    \includegraphics[width=0.49\textwidth]{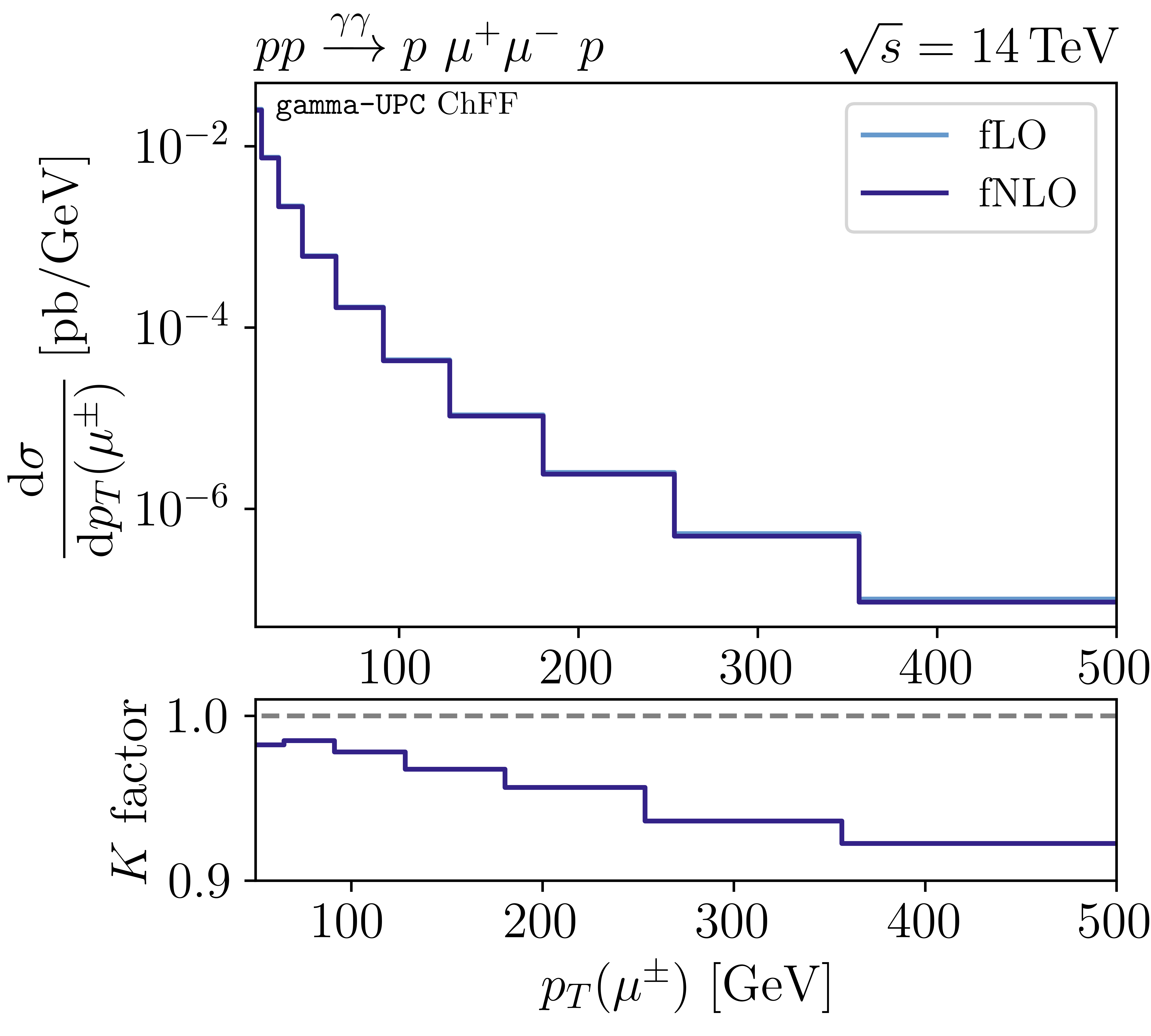}
    \caption{Differential cross sections for dimuon production as a function of the invariant mass, $\mathrm{d}\sigma/\mathrm{d}m_{\mumu}$ (left), and the transverse momentum of $\mu^\pm$, $\mathrm{d}\sigma/\mathrm{d}p_{T}(\mu^\pm)$ (right), in \pp\ UPCs at $\sqrt{s}=14$ TeV, within the fiducial cuts defined in section \ref{sec:setup}. The lower panels display the $K$ factors.}
    \label{fig:aa2mumu}
\end{figure}

For the differential cross sections, we present the invariant mass $m_{\mumu}$ (left) and transverse momentum $p_{T}(\mu^\pm)$ (right) distributions at 14 TeV in \pp\ collisions in figure \ref{fig:aa2mumu}. The differential $K$ factors are shown in the lower panels. The NLO EW corrections are negative, and their relative contributions with respect to LO generally increase with $m_{\mumu}$ and $p_T(\mu^{\pm})$, except when $m_{\mumu}$ is around the $W^\pm$ pair threshold, $2M_W\simeq 160$ GeV. At this point, we suspect that the bump in the $K$ factor arises due to the presence of an anomalous threshold~\cite{Boudjema:2008zn,Passarino:2018wix} in one-loop Feynman diagrams. The largest NLO EW correction in the figure occurs in the tail of the $p_T(\mu^{\pm})$ distribution, reaching approximately $-8\%$.

If electrons and positrons are treated as dressed leptons, the results presented for dimuon production should also apply to the Breit–Wheeler process, $\gaga\to \epem$, since the lepton mass effects are negligible. However, we do not consider the process $\gaga\to \epem$ in this study.

\subsection{$\gaga\to \tautau$\label{sec:ditau}}

The ditau production process in photon fusion is of particular interest because it directly probes the $\gamma\tautau$ vertex and is thus sensitive to the anomalous electromagnetic couplings of the $\tau$ lepton. The $\gaga\to\tautau$ process has long been proposed as a probe for the anomalous magnetic moment $a_\tau$ and the electric dipole moment $d_\tau$ of the tau lepton at both lepton~\cite{Cornet:1995pw} and hadron~\cite{delAguila:1991rm,Atag:2010ja} colliders. This process has also been measured in various experiments, including those conducted at the Large Electron-Positron (LEP) collider~\cite{OPAL:1993law,L3:1997acq,DELPHI:2003nah} and in UPCs at the LHC~\cite{ATLAS:2022ryk,CMS:2022arf,CMS:2024qjo}. Notably, the best constraint on $a_\tau$, $a_\tau=0.0009\left(^{+0.0016}_{-0.0015}\right)_{\mathrm{syst}}\left(^{+0.0028}_{-0.0027}\right)_{\mathrm{stat}}$ at the $68\%$ confidence level, has been obtained by the CMS collaboration in \pp\ UPCs at 13 TeV~\cite{CMS:2024qjo}. Since the leading contribution to $a_\tau$ in the SM, originally derived by Schwinger~\cite{Schwinger:1948iu} and universal for all leptons, enters the $\gaga\to\tautau$ cross section through NLO radiative corrections, it is essential to compute NLO EW corrections for this process in UPCs. NLO QED corrections have been reported in ref.~\cite{Shao:2024dmk}, while NLO EW corrections in \PbPb\ UPC have been calculated in ref.~\cite{Jiang:2024dhf}. A comparison between our automated calculations and the results in ref.~\cite{Jiang:2024dhf} is therefore of particular interest. 

To facilitate this comparison, we retain the masses of the charm quark, tau lepton, and bottom quark as given in table \ref{tab:EWparaminput} and generate the process using the following commands:
\vskip 0.25truecm
%\noindent
%~~\prompt\ {\tt ~set~complex\_mass\_scheme~true}

\noindent
~~\prompt\ {\tt ~import~model~loop\_qcd\_qed\_sm\_Gmu\_3FS-with\_b\_mass\_tau\_mass\_c\_mass\_a0}

\noindent
~~\prompt\ {\tt ~generate \taggeda\ \taggeda\ > ta+ ta- [QED]}

\vskip 0.25truecm
\noindent
As in the dimuon case discussed in section \ref{sec:dimuon}, we have $k_0=2, c_s(k_0)=0, c(k_0)=2$, and $\Delta(k_0)=0$, with $\Sigma_{\mathrm{NLO}_1}=0$ for $\gaga\to\tautau$. 

\begin{table}[t]
\centering
\tabcolsep=3.5mm
\vspace{0.2cm}
\begin{tabular}{l|c|c|c} \hline
Process: $\gaga\to \tautau$ & \multicolumn{3}{c}{\gammaUPC+\mgshort} \\
Colliding system, \cm\ energy & $\sigma_{\mathrm{LO}}$ & $\sigma_{\mathrm{NLO}\ \mathrm{EW}}$ & $\sigma_{\mathrm{NLO}\ \mathrm{EW}}/\sigma_{\mathrm{LO}}$ \\ \hline
\pp\ at 13~TeV &  $190.57(6)$ pb &  $192.30(6)$ pb & 1.009 \\
\pp\ at 13.6~TeV &  $194.93(7)$ pb &  $196.68(7)$ pb & 1.009 \\
\pp\ at 14~TeV &  $197.67(6)$ pb &  $199.45(7)$ pb & 1.009 \\
\pPb\ at 8.8~TeV &  $564.3(2)$ nb &  $569.6(2)$ nb & 1.009 \\
\PbPb\ at 5.52~TeV &  $1.1628(4)$ mb &  $1.1742(4)$ mb & 1.010 \\
\hline
\pp\ at 100~TeV &  $459.5(2)$ pb &  $463.6(2)$ pb & 1.009 \\
\pPb\ at 62.8~TeV &  $1.6578(6)$ $\mu$b &  $1.6727(6)$ $\mu$b & 1.009 \\
\PbPb\ at 39.4~TeV &  $5.117(2)$ mb &  $5.164(2)$ mb & 1.009 \\
\hline
\end{tabular}
\caption{The total inclusive ditau cross sections at LO and NLO accuracy at LHC and FCC-hh energies, computed using the ChFF $\gamma$ fluxes available in \gammaUPC. The last column also shows the corresponding $K$ factors. Numbers in parentheses indicate statistical errors from numerical integrations, which may affect the last decimal place.\label{tab:ditauxs}}
\end{table}

The integrated inclusive cross sections without any kinematic cuts, along with the corresponding $K$ factors, are reported in table \ref{tab:ditauxs} for LHC and FCC-hh energies using the ChFF $\gamma$ fluxes. Unlike the findings in ref.~\cite{Jiang:2024dhf}, our results show that the NLO EW corrections lead to a $1\%$ enhancement of the LO cross sections, which is in fact quite close to the NLO QED corrections reported in ref.~\cite{Shao:2024dmk} (see the first row of table 3 therein). This agreement is expected, as weak corrections should be suppressed by a factor of $\mathcal{O}\left(M_\tau^2/M_W^2\right)$ relative to QED corrections. In contrast to ref.~\cite{Jiang:2024dhf}, we do not observe the reported $-4\%$ weak corrections. Given that our framework has been extensively tested, we are confident in the accuracy of our results.~\footnote{After the submission of our paper to arXiv, a preprint~\cite{Dittmaier:2025ikh} appeared, presenting NLO EW corrections to the $2\to 6$ process $\gaga\to\tautau\to e^+\mu^-\bar{\nu}_\tau \nu_\tau \bar{\nu}_\mu \nu_e$ in the (improved) narrow-width approximation. According to table 1 of that paper, the main discrepancy between our results and those of ref.~\cite{Jiang:2024dhf} originates from the use of different EW renormalization schemes. In our work, we adopt a hybrid $\alpha$ renormalization scheme, which is close to the $\alpha(0)$ scheme used in ref.~\cite{Dittmaier:2025ikh}, while ref.~\cite{Jiang:2024dhf} employs the $G_\mu$ scheme. As discussed in section \ref{sec:theory}, however, our scheme provides a more appropriate choice than the $G_\mu$ scheme for this process.}

\begin{figure}
    \includegraphics[width=0.49\textwidth]{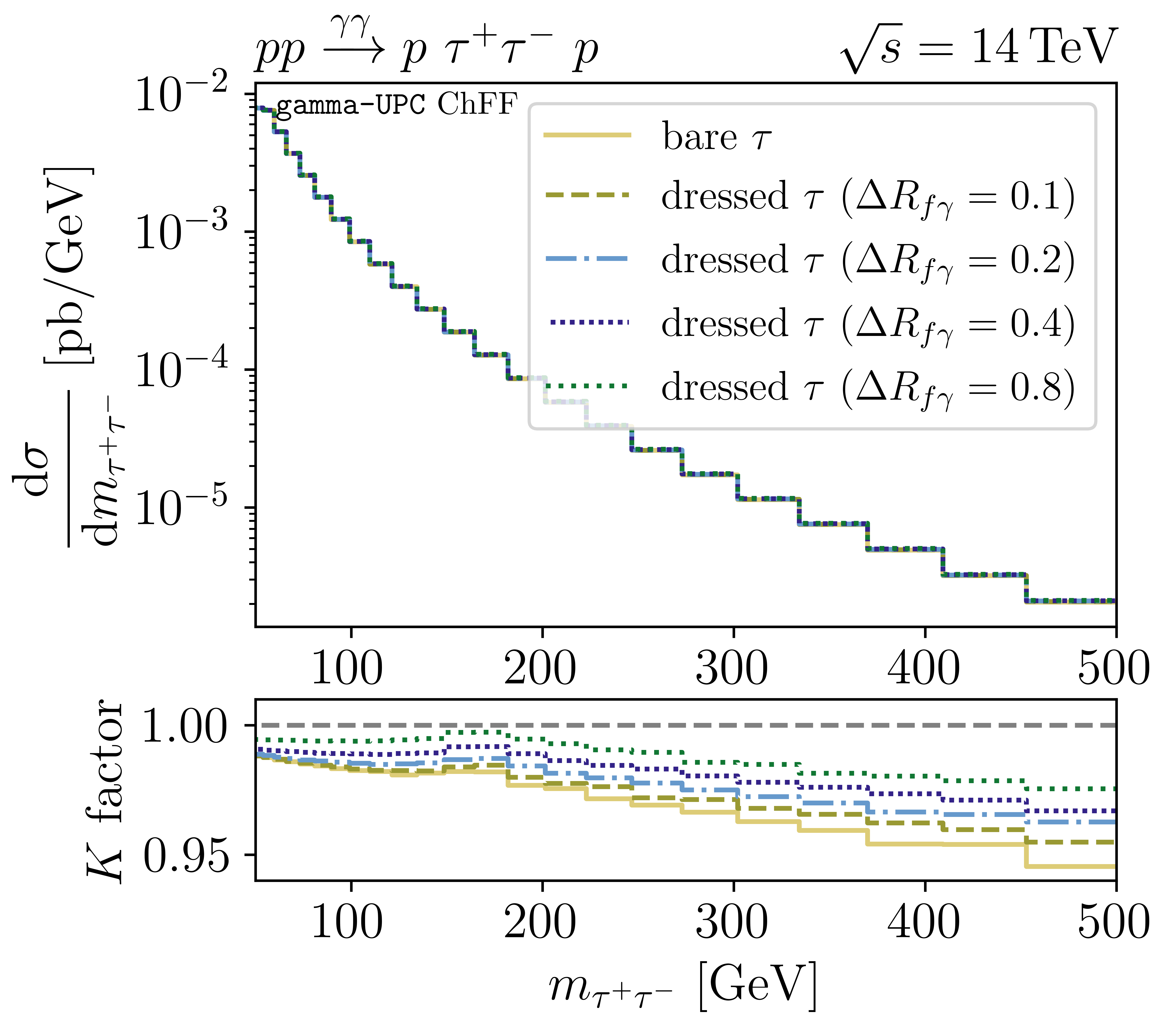}
    \includegraphics[width=0.49\textwidth]{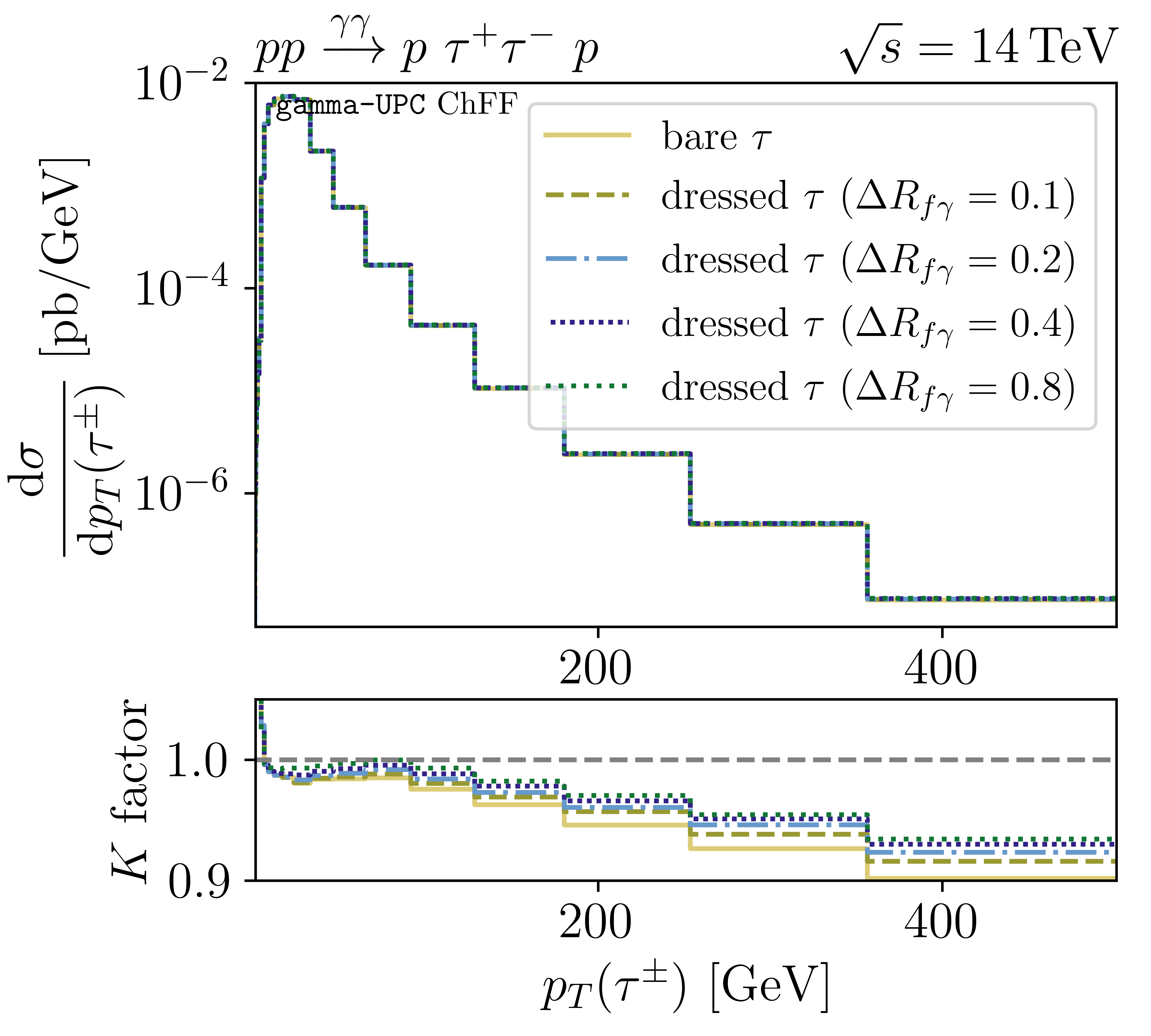}
    \caption{Differential cross sections for ditau production as functions of the invariant mass, $\mathrm{d}\sigma/\mathrm{d}m_{\tautau}$ (left), and the transverse momentum of $\tau^\pm$, $\mathrm{d}\sigma/\mathrm{d}p_{T}(\tau^\pm)$ (right), in \pp\ UPCs at $\sqrt{s}=14$ TeV, with the cuts $m_{\tautau}>50$ GeV and $|\eta(\tau^\pm)|<2.5$. The lower panels show the $K$ factors. The distributions for the bare tau are compared to those for the dressed tau lepton, defined with four different values of $\Delta R_{f\gamma}$.}
    \label{fig:aa2tautau}
\end{figure}

With a nonzero tau mass, we can compare results computed with bare leptons to those with dressed leptons. Adopting similar fiducial cuts ($m_{\tautau}>50$ GeV and $|\eta(\tau^\pm)|<2.5$) as in the CMS measurement~\cite{CMS:2024qjo}, the differential cross sections as functions of the invariant mass $m_{\tautau}$ (left) and transverse momentum $p_T(\tau^\pm)$ (right) are shown in figure~\ref{fig:aa2tautau}. The behavior of the $K$ factors is qualitatively similar to the dimuon case: the NLO EW corrections are negative, and their relative contributions increase with $m_{\tautau}$ and $p_T(\tau^\pm)$. The anomalous threshold at $m_{\tautau}\sim 2M_W$ is clearly visible. Additionally, we compare the $K$ factors obtained using the bare tau lepton and the dressed lepton in the same figure. The significance of NLO EW corrections decreases when transitioning from the bare tau (equivalent to using $\Delta R_{f\gamma}=0$ in the photon recombination procedure) to the dressed tau with larger $\Delta R_{f\gamma}$. This aligns with the general expectation that (quasi-)collinear logarithms should cancel out in ``dressed" leptons, as discussed in section \ref{sec:dimuon}. Figure \ref{fig:aa2tautau} provides a quantitative example of this effect.

\subsection{$\gaga\to \WpWm$\label{sec:WW}}

The two-photon production of a $W$ boson pair is recognized as an ideal probe of anomalous quartic gauge couplings~\cite{Pierzchala:2008xc,Maniatis:2008zz,deFavereaudeJeneret:2009db,Chapon:2009hh,Bailey:2022wqy,Shao:2022cly}, which can originate from UV-complete BSM theories. Consequently, this process has been measured at both the Tevatron~\cite{D0:2013rce} and the LHC~\cite{CMS:2013hdf,CMS:2016rtz,ATLAS:2016lse,ATLAS:2020iwi,CMS:2022dmc}. The exclusive $\WpWm$ production cross section, after excluding proton dissociative contributions, has been determined by the ATLAS~\cite{ATLAS:2016lse} and CMS~\cite{CMS:2016rtz} experiments in 8 TeV \pp\ collisions as
\begin{eqnarray}
\sigma_{\mathrm{ATLAS}}(\gaga\to \WpWm\to e^\pm \mu^\mp X)&=&2.09\pm 1.12~\mathrm{fb}\,,\\
\sigma_{\mathrm{CMS}}(\gaga\to \WpWm\to e^\pm \mu^\mp X)&=&2.4^{+1.5}_{-1.2}~\mathrm{fb}\,.
\end{eqnarray}
These results are in agreement with our state-of-the-art SM prediction, computed using \mgshort\ with the ChFF photon flux implemented in \gammaUPC,
\begin{equation}
\sigma_{\mathrm{NLO~EW}}(\gaga\to \WpWm)\mathrm{Br}(\WpWm\to e^\pm \mu^\mp X)=0.989~\mathrm{fb}\,,
\end{equation}
where the branching ratio for $\WpWm$ decaying into $e^\pm \mu^\mp X$ is $\mathrm{Br}(\WpWm\to e^\pm \mu^\mp X)=3.23\%$, accounting for the leptonic decay channels of the $\tau$ lepton. 

\begin{table}[t]
\centering
\tabcolsep=3.5mm
\vspace{0.2cm}
\begin{tabular}{l|c|c|c} \hline
Process: $\gaga\to \WpWm$ & \multicolumn{3}{c}{\gammaUPC+\mgshort} \\
Colliding system, \cm\ energy & $\sigma_{\mathrm{LO}}$ & $\sigma_{\mathrm{NLO}\ \mathrm{EW}}$ & $\sigma_{\mathrm{NLO}\ \mathrm{EW}}/\sigma_{\mathrm{LO}}$ \\ \hline
\pp\ at 8~TeV &  $30.41(1)$ fb &  $30.62(1)$ fb & 1.007 \\
\pp\ at 13~TeV &  $63.06(2)$ fb &  $63.19(2)$ fb & 1.002 \\
\pp\ at 13.6~TeV &  $67.16(2)$ fb &  $67.27(2)$ fb & 1.002 \\
\pp\ at 14~TeV &  $69.82(2)$ fb &  $69.91(2)$ fb & 1.001 \\
\pPb\ at 8.8~TeV &  $30.47(1)$ pb &  $30.98(1)$ pb & 1.017 \\
\PbPb\ at 5.52~TeV &  $325.3(1)$ pb &  $336.3(1)$ pb & 1.034 \\
\hline
\pp\ at 100~TeV &  $559.1(2)$ fb &  $548.6(2)$ fb & 0.981 \\
\pPb\ at 62.8~TeV &  $824.4(2)$ pb &  $821.6(3)$ pb & 0.997 \\
\PbPb\ at 39.4~TeV &  $490.5(2)$ nb &  $497.3(2)$ nb & 1.014 \\
\hline
\end{tabular}
\caption{The total inclusive cross sections for $\gaga\to \WpWm$ at LO and NLO accuracy at LHC and FCC-hh energies, computed using the ChFF $\gamma$ fluxes available in \gammaUPC. The last column also shows the corresponding $K$ factors. Numbers in parentheses indicate statistical errors from numerical integrations, which may affect the last decimal place.\label{tab:WWxs}}
\end{table}

In order to calculate the NLO EW cross sections, we execute the following commands within the \mgshort\ framework, using the five-flavor scheme with $M_c=M_\tau=M_b=0$:
\vskip 0.25truecm
\noindent
~~\prompt\ {\tt ~import~model~loop\_qcd\_qed\_sm\_Gmu\_3FS-a0}

\noindent
~~\prompt\ {\tt ~generate \taggeda\ \taggeda\ > w+ w- [QED]}

\vskip 0.25truecm
\noindent
The integrated LO and NLO EW cross sections for \pp, \pPb, and \PbPb\ collisions are presented in table \ref{tab:WWxs}. The $K$ factors in the last column indicate that accidental cancellations occur in the radiative corrections for \pp\ collisions at around 13-14 TeV, while such cancellations are not observed in heavy-ion collisions due to the significantly different photon fluxes of lead and proton. At NLO EW, the corrections to the total cross sections of $\gaga\to\WpWm$ remain at the percent level.

\subsection{$\gaga\to \ttbar$\label{sec:ttbar}}

The photon-fusion production of a pair of top quarks in UPCs with protons and ions at hadron colliders provide complementary information to processes such as the inclusive reaction $pp\to \ttbar \gamma_{\mathrm{iso}}+X$~\cite{Pagani:2021iwa} for testing the interaction of top quarks and photons, as well as for determining top-quark properties, such as the electric charge, electromagnetic dipole moments of the top quark, and any potential BSM enhancement. A first measurement of this process was carried out by the CMS-TOTEM collaboration~\cite{CMS:2023naq} at 13 TeV in \pp\ collisions at the LHC, by tagging two forward intact protons with the precision proton spectrometer. Although only an upper bound on the production cross section has been set so far, the observation of the central exclusive production of top-quark pairs is expected to be possible at the high-luminosity LHC~\cite{CMS:2021ncv,Goncalves:2020saa,Martins:2022dfg}. The cross sections for $\gaga\to\ttbar$ in UPCs are known at NLO QCD accuracy~\cite{Shao:2022cly} based on \gammaUPC\ and a custom code generated by \mgshort. In our automated computational framework, we can easily include NLO EW corrections with the following generation commands:
\vskip 0.25truecm

\noindent
~~\prompt\ {\tt ~import~model~loop\_qcd\_qed\_sm\_Gmu\_3FS-a0}

\noindent
~~\prompt\ {\tt ~generate \taggeda\ \taggeda\ > t t\~{} [QCD QED]}

\vskip 0.25truecm
\noindent The process has $k_0=2$, $c_s(k_0)=0$, $c(k_0)=2$, and $\Delta(k_0)=0$. There is one LO term $\Sigma_{\mathrm{LO}_1}$ at $\mathcal{O}(\alpha^2)$ and two NLO terms, $\Sigma_{\mathrm{NLO}_1}$ at $\mathcal{O}(\alpha^2\alpha_s)$ and $\Sigma_{\mathrm{NLO}_2}$ at $\mathcal{O}(\alpha^3)$.

\begin{table}[htpb!]
\centering
\tabcolsep=3.5mm
%\caption{CAPTION}
%\label{tab:LABEL}
\vspace{0.2cm}
\begin{tabular}{l|c|c|c} \hline
Process: $\gaga\to \ttbar$ & \multicolumn{3}{c}{\gammaUPC+\mgshort} \\
Colliding system, \cm\ energy & $\sigma_{\mathrm{LO}}$ & $\sigma_{\mathrm{NLO}\ \mathrm{QCD}}$ & $\sigma_{\mathrm{NLO}\ \mathrm{QCD+EW}}$ \\ \hline
\pp\ at 13~TeV &  $212.40(6)$ ab &  $256.43(9)^{+4.5}_{-3.7}$ ab & $244.8(1)^{+4.5}_{-3.7}$ ab\\
\pp\ at 13.6~TeV &  $228.53(6)$ ab &  $275.5(1)^{+4.8}_{-4.0}$ ab & $263.1(1)^{+4.8}_{-4.0}$ ab \\
\pp\ at 14~TeV &  $239.58(7)$ ab &  $288.7(1)^{+5.0}_{-4.2}$ ab & $275.5(1)^{+5.0}_{-4.2}$ ab \\
\pPb\ at 8.8~TeV &  $46.89(1)$ fb &  $59.87(2)^{+1.3}_{-1.1}$ fb & $57.32(2)^{+1.3}_{-1.1}$ fb \\
\PbPb\ at 5.52~TeV &  $30.64(1)$ fb &  $39.08(1)^{+0.87}_{-0.72}$ fb & $37.43(1)^{+0.87}_{-0.72}$ fb \\
\pp\ at 100~TeV & $2.3080(2)$ fb &  $2.7111(2)^{+0.041}_{-0.034}$ fb & $2.5816(2)^{+0.041}_{-0.034}$ fb \\
\pPb\ at 62.8~TeV &  $3.0742(2)$ pb &  $3.6721(3)^{+0.061}_{-0.050}$ pb & $3.5045(3)^{+0.061}_{-0.050}$ pb \\
\PbPb\ at 39.4~TeV &  $0.9583(1)$ nb &  $1.2062(2)^{+0.026}_{-0.021}$ nb & $1.1545(2)^{+0.026}_{-0.021}$ nb \\
\hline
 $K$ factor & & $\sigma_{\mathrm{NLO}\ \mathrm{QCD}}/\sigma_{\mathrm{LO}}$ & $\sigma_{\mathrm{NLO}\ \mathrm{QCD+EW}}/\sigma_{\mathrm{LO}}$ \\ \hline
\pp\ at 13~TeV & & 1.207 & 1.153 \\
\pp\ at 13.6~TeV & & 1.205 & 1.151 \\
\pp\ at 14~TeV & & 1.205 & 1.151 \\
\pPb\ at 8.8~TeV & & 1.277 & 1.222 \\
\PbPb\ at 5.52~TeV & & 1.276 & 1.222 \\
\pp\ at 100~TeV & &  1.175 & 1.119 \\
\pPb\ at 62.8~TeV &  &  1.194 & 1.140 \\
\PbPb\ at 39.4~TeV &  &  1.259 & 1.205 \\
\hline
\end{tabular}
\caption{The inclusive cross sections for $\gaga\to \ttbar$ at LO, NLO QCD, and NLO QCD+EW accuracy at LHC and FCC-hh energies, computed using the ChFF $\gamma$ fluxes available in \gammaUPC. The quoted uncertainties in the absolute cross sections arise
from variations of the renormalization scale by a factor of $2$. Numbers in parentheses indicate statistical errors from numerical integrations, which may affect the last decimal place.\label{tab:ttbarxs}}
\end{table}

The inclusive cross sections at the LHC and FCC-hh are reported in table \ref{tab:ttbarxs}, along with the $K$ factors. These are presented at LO, NLO QCD, and NLO QCD+EW accuracies, respectively.\footnote{For LO and NLO QCD cross sections, they agree with ref.~\cite{Shao:2022cly} using the same setup, except for those evaluated at $\sqrtsnn=5.52$ TeV in \PbPb\ collisions, due to numerical instabilities in evaluating the effective photon-photon luminosity of \PbPb\ at high $x$ in the initial version of \gammaUPC.} The lower cross section in \PbPb\ at $\sqrtsnn=5.52$ TeV compared to \pPb\ at $\sqrtsnn=8.8$ TeV can be easily explained by their effective photon-photon luminosities in the high-mass region (cf. figure 2 in ref.~\cite{dEnterria:2025ecx}). Our study shows that NLO QCD corrections enhance the LO cross sections by $20\%$ ($17\%$) in \pp\ and $28\%$ ($26\%$) in \PbPb\ collisions at the LHC (FCC-hh). On the other hand, NLO EW calculations result in a $-5.5\%$ reduction of the LO cross sections. Our state-of-the-art predictions incorporate both QCD and EW radiative corrections. The (renormalization) scale uncertainties due to the strong interaction are around $\pm2\%$. However, using scale variation to estimate missing higher-order corrections likely underestimates the size of unknown next-to-NLO (NNLO) QCD corrections. Our result emphasizes the need to include NLO corrections for accurate calculations of cross sections for top-quark pairs or any hadronic final states in UPCs. Given the cross sections and integrated luminosities (cf. table II in ref.~\cite{Shao:2022cly}),  this process can only be observed in \pp\ collisions with forward proton tagging at the LHC.

\begin{figure}
    \includegraphics[width=0.49\textwidth]{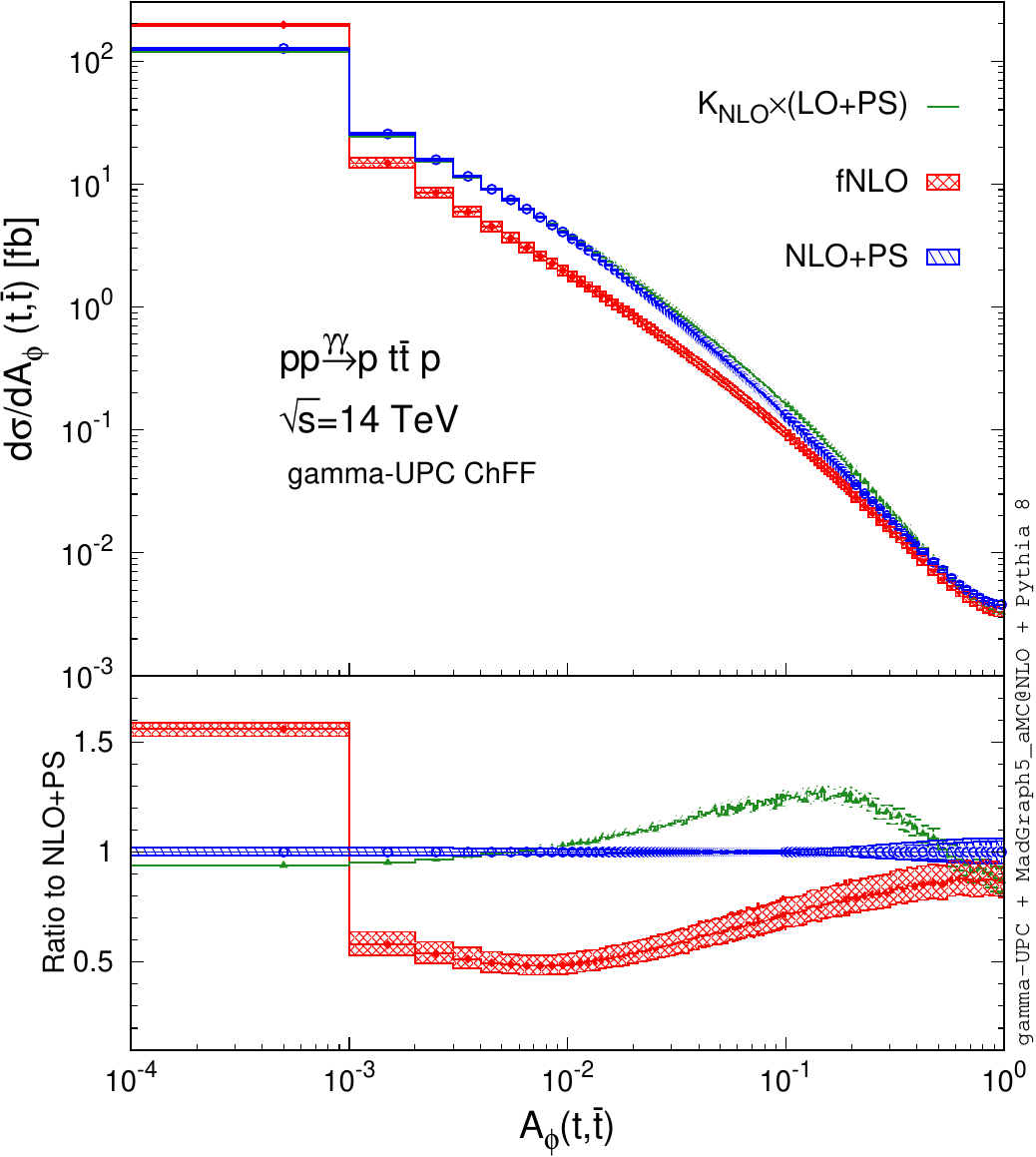}
    \includegraphics[width=0.49\textwidth]{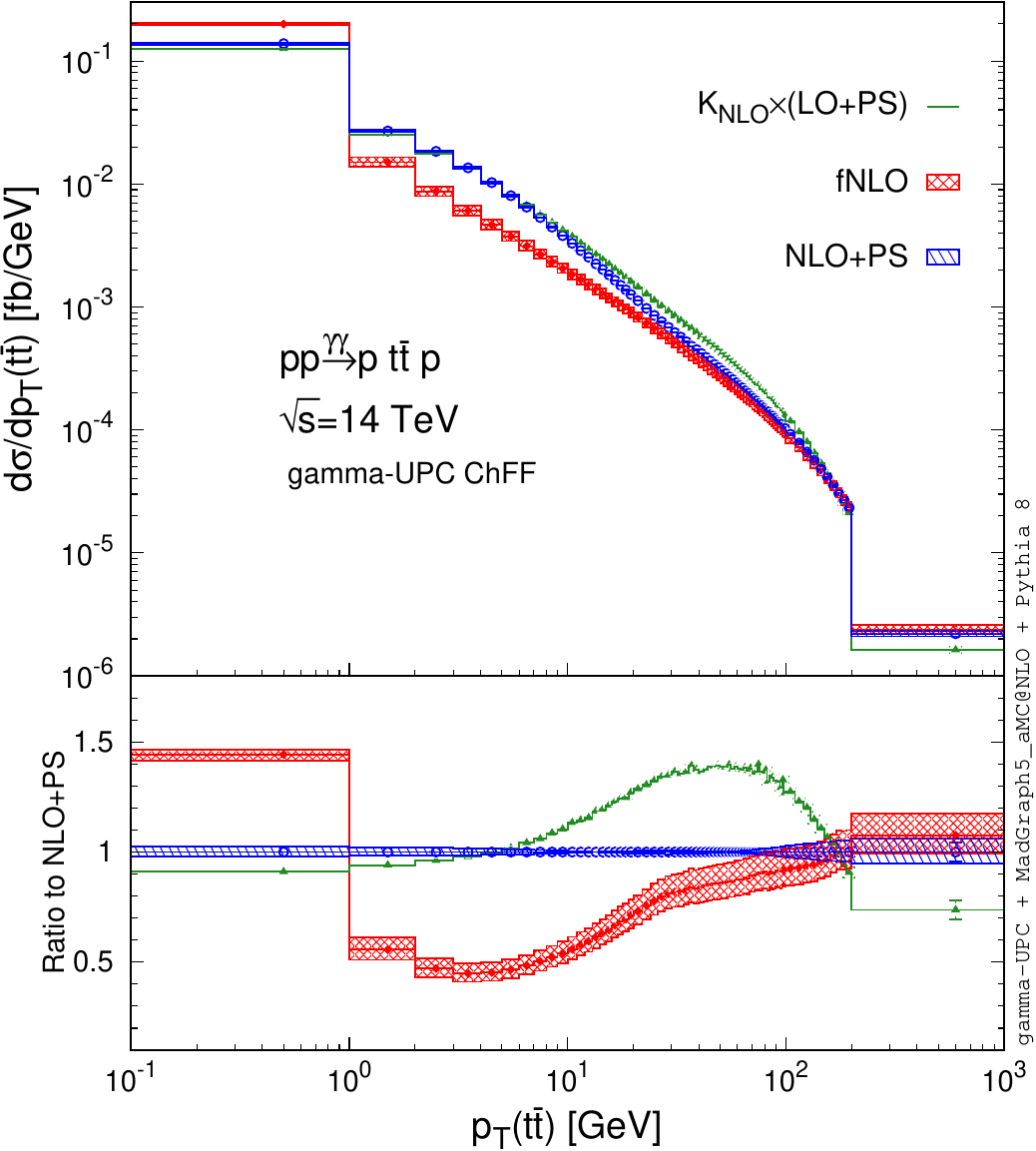}
    \caption{Differential distributions of the top-quark pair acoplanarity $\mathrm{d}\sigma/\mathrm{d}A_{\phi}(t,\bar{t})$ (left) and transverse momentum $\mathrm{d}\sigma/\mathrm{d}p_T(t\bar{t})$ (right) in \pp\ UPCs at $\sqrt{s}=14$ TeV, computed using the ChFF $\gamma$ fluxes available in \gammaUPC\ at LO+PS (green histograms), fNLO (red histograms), and NLO+PS (blue histograms) accuracy. Here, we refer to fNLO and NLO+PS as NLO QCD only. The LO+PS histograms have been rescaled by an overall $K$ factor, $K_{\mathrm{NLO}}\equiv\sigma_{\mathrm{NLO}\ \mathrm{QCD}}/\sigma_{\mathrm{LO}}\approx 1.21$. The lower panels display the ratios relative to the central values of NLO+PS. The hatched bands in fNLO and NLO+PS indicate renormalization scale uncertainties, while the error bars represent statistical uncertainties. The lower ranges of the first (left-most) bins in both plots should extend to zero.}
    \label{fig:aa2ttbarNLOPS}
\end{figure}

As indicated in table \ref{tab:allowedinMG}, if we consider only NLO QCD corrections for processes without jets, the program also allows for NLO+PS simulations using the \mcatnlo\ method. Figure \ref{fig:aa2ttbarNLOPS} presents two observables in \pp\ UPCs at $\sqrt{s}=14$ TeV: the top-quark pair acoplanarity $A_{\phi}(t,\bar{t})\equiv 1-|\Delta \phi(t,\bar{t})|/\pi$ (left) and transverse momentum $p_T(t\bar{t})$ (right), both of which are sensitive to the PS effect. We utilize \gammaUPC\ to model the ChFF photon flux and \pythia8.235~\cite{Sjostrand:2014zea} for final-state parton showering and hadronization. Figure \ref{fig:aa2ttbarNLOPS} compares NLO+PS with both the fixed-order NLO QCD computation (fNLO, red histograms) and LO+PS (green histograms). The LO+PS histograms have been multiplied by an overall $K$ factor, $K_{\mathrm{NLO}}\equiv\sigma_{\mathrm{NLO}\ \mathrm{QCD}}/\sigma_{\mathrm{LO}}\approx 1.21$. The lower panels display the ratio of each result to the central values of the NLO+PS histograms. The hatched bands for fNLO and NLO+PS represent renormalization scale uncertainties, and the error bars represent statistical errors, where we have generated 20M and 40M events for LO+PS and NLO+PS, respectively. To improve visibility, we truncate the lower ranges of the first bins in figure \ref{fig:aa2ttbarNLOPS} to nonzero values, although they should extend to zero. In the limit of vanishing $A_{\phi}(t,\bar{t})$ or $p_T(t\bar{t})$, fNLO is divergent. In contrast, PS resums logarithmically enhanced terms, leading to more realistic LO+PS and NLO+PS predictions in this regime. As a result, LO+PS and NLO+PS predictions are similar to each other and significantly lower than the fNLO prediction. Additionally, in the tails of the two distributions, NLO+PS should asymptotically approach fNLO since these regions are dominated by hard radiation. The comparisons among the three results for the two distributions indeed follow the expected pattern. In the transition region between the two asymptotic limits, rescaled LO+PS and fNLO produce harder and softer spectra than NLO+PS, respectively. LO+PS maximally overestimates the NLO+PS spectra by 30-50\%, while fNLO underestimates them by at most a factor of $2$. We note that when $A_{\phi}(t,\bar{t})>0$ or $p_{T}(t\bar{t})>0$, fNLO is actually at LO accuracy, with contributions only from real emission diagrams. Consequently, the fractional renomormalization scale dependence is substantial, around $\pm10\%$. In contrast, the low and intermediate regions of the NLO+PS histograms are mainly populated through the showering of $\mathbb{S}$ events in the \mcatnlo\ formalism, whose weights receive both $\mathcal{O}(\alpha_s^0)$ and $\mathcal{O}(\alpha_s)$ contributions. Hence, the relative scale uncertainties of NLO+PS in figure \ref{fig:aa2ttbarNLOPS} are significantly smaller than those of fNLO in these regions. In the tails, the scale dependence of NLO+PS grows as expected. Finally, we wish to comment that we are working within the strictly collinear factorization framework and have ignored any (small) virtuality dependence of the initial-state photons. The inclusion of intrinsic photon virtuality may modify the low $A_{\phi}(t,\bar{t})$ and low $p_T(t\bar{t})$ distributions, analogous to the dimuon case shown in figure 3 of ref.~\cite{Shao:2024dmk}. However, we anticipate that this effect is negligible for $A_{\phi}(t,\bar{t})>0.04$ or $p_T(t\bar{t})>1$ GeV.

\subsection{$\gaga\to \ttbar\gamma_{\mathrm{iso}}$\label{sec:ttbargamma}}

Although it is likely hopeless to observe experimentally at the LHC, we consider the photon-fusion production of a top quark pair in association with an isolated photon as a showcase to demonstrate that our implementation is compatible with both initial coherent photons and final tagged photons, as enabled in ref.~\cite{Pagani:2021iwa}. This process can be generated via:
\vskip 0.25truecm
\noindent
~~\prompt\ {\tt ~import~model~loop\_qcd\_qed\_sm\_Gmu\_3FS-a0}

\noindent
~~\prompt\ {\tt ~generate \taggeda\ \taggeda\ > t t\~{} \taggeda\ [QCD QED]}

\vskip 0.25truecm
\noindent
Similar to $\gaga\to\ttbar$, both NLO QCD and EW corrections are nonzero.

\begin{table}[htpb!]
\centering
\tabcolsep=3.5mm
%\caption{CAPTION}
%\label{tab:LABEL}
\vspace{0.2cm}
\begin{tabular}{l|c|c|c} \hline
Process: $\gaga\to \ttbar\gamma_\mathrm{iso}$ & \multicolumn{3}{c}{\gammaUPC+\mgshort} \\
Colliding system, \cm\ energy & $\sigma_{\mathrm{LO}}$ & $\sigma_{\mathrm{NLO}\ \mathrm{QCD}}$ & $\sigma_{\mathrm{NLO}\ \mathrm{QCD+EW}}$ \\ \hline
\pp\ at 13~TeV & $0.4980(2)$ ab & $0.5325(2)^{+0.0034}_{-0.0028}$ ab & $0.4997(1)^{+0.0034}_{-0.0028}$ ab \\
\pp\ at 13.6~TeV & $0.5439(2)$ ab & $0.5809(2)^{+0.0037}_{-0.0030}$ ab & $0.5448(2)^{+0.0037}_{-0.0030}$ ab \\
\pp\ at 14~TeV & $0.5753(2)$ ab & $0.6144(2)^{+0.0039}_{-0.0032}$ ab & $0.5757(2)^{+0.0038}_{-0.0032}$ ab \\
\pPb\ at 8.8~TeV &  $44.62(1)$ ab & $50.36(2)^{+0.58}_{-0.48}$ ab & $47.78(2)^{+0.57}_{-0.48}$ ab \\
\PbPb\ at 5.52~TeV &  $35.09(1)$ ab & $38.27(1)^{+0.32}_{-0.26}$ ab & $36.18(1)^{+0.32}_{-0.26}$ ab \\
\pp\ at 100~TeV & $7.622(2)$ ab & $7.988(3)^{+0.034}_{-0.029}$ ab & $7.355(3)^{+0.034}_{-0.029}$ ab \\
\pPb\ at 62.8~TeV & $6.587(2)$ fb & $6.983(3)^{+0.039}_{-0.032}$ fb & $6.509(3)^{+0.038}_{-0.032}$ fb \\
\PbPb\ at 39.4~TeV & $1.2875(4)$ pb & $1.4364(5)^{+0.015}_{-0.012}$ pb & $1.3615(5)^{+0.015}_{-0.012}$ pb \\
\hline
 $K$ factor & & $\sigma_{\mathrm{NLO}\ \mathrm{QCD}}/\sigma_{\mathrm{LO}}$ & $\sigma_{\mathrm{NLO}\ \mathrm{QCD+EW}}/\sigma_{\mathrm{LO}}$ \\ \hline
\pp\ at 13~TeV & & 1.069 & 1.003 \\
\pp\ at 13.6~TeV & & 1.068 & 1.002 \\
\pp\ at 14~TeV & & 1.068 & 1.001 \\
\pPb\ at 8.8~TeV & & 1.129 & 1.071 \\
\PbPb\ at 5.52~TeV & & 1.091 & 1.031 \\
\pp\ at 100~TeV & & 1.048 & 0.965 \\
\pPb\ at 62.8~TeV & & 1.060 & 0.989 \\
\PbPb\ at 39.4~TeV & & 1.116 & 1.057 \\ \hline
\end{tabular}
\caption{The integrated fiducial cross sections (within the fiducial cuts defined in section \ref{sec:setup}) for $\gaga\to \ttbar\gamma_{\mathrm{iso}}$ at LO, NLO QCD, and NLO QCD+EW accuracy at LHC and FCC energies, computed using the ChFF $\gamma$ fluxes available in \gammaUPC. The quoted uncertainties in the absolute cross sections arise
from variations of the renormalization scale by a factor of $2$. Numbers in parentheses indicate statistical errors from numerical integrations, which may affect the last decimal place.\label{tab:ttbargammaxs}}
\end{table}

The phase-space integrated fiducial cross sections are shown in table \ref{tab:ttbargammaxs}, where the definition of the isolated photons is provided in section \ref{sec:setup}. NLO QCD corrections enhance the LO cross sections by $7\%$, $13\%$, and $9\%$ in \pp, \pPb, and \PbPb\ collisions at the LHC, respectively. At the FCC-hh, the QCD correction amounts change to $5\%$, $6\%$, and $12\%$. NLO EW corrections reduce the cross sections by around 6-8\%, with mildly dependence on the colliding system and \cm\ energy. The two types of quantum corrections are accidentally canceled in the \pp\ collision mode at the LHC. Unfortunately,  we are still unable to explain why these cancellations occur.

\subsection{$\gaga \to \ttbar j$\label{sec:ttbarjet}}
In this section, we take the process $\gaga\to \ttbar j$ as a showcase for computing NLO cross sections of processes with jets in the final state. From na\"ive coupling power counting, this process is suppressed by a factor of $\alpha_s\approx 0.1$ relative to $\gaga\to \ttbar$, which was already discussed in section \ref{sec:ttbar}. Although rare, this process remains measurable at the high-luminosity LHC in \pp\ collisions. 

To compute NLO cross sections, this process can be generated by executing the following commands:
\vskip 0.25truecm
\noindent
~~\prompt\ {\tt ~set~complex\_mass\_scheme~True}

\noindent
~~\prompt\ {\tt ~import~model~loop\_qcd\_qed\_sm\_Gmu\_3FS-a0}

\noindent
~~\prompt\ {\tt ~define j = u c d s b u\~{} c\~{} d\~{} s\~{} b\~{} e+ mu+ ta+ e- mu- ta- a g}

\noindent
~~\prompt\ {\tt ~generate \taggeda\ \taggeda\ > t t\~{} j aS=1 aEW=3 [QCD QED]}

\vskip 0.25truecm
\noindent
Using the notation introduced in section \ref{sec:theory}, we deduce the following parameters:
\begin{align}
k_0=&3, \quad c_s(k_0)=0, \quad c(k_0)=2, \quad \Delta(k_0)=1\,.
\end{align}
This implies that the process consists of two LO contributions:
\begin{equation}
\Sigma_{\mathrm{LO}_1}~(\mathcal{O}(\alpha_s\alpha^2)),\quad \Sigma_{\mathrm{LO}_2}~(\mathcal{O}(\alpha^3))\,,
\end{equation}
and three NLO contributions:
\begin{equation}
\Sigma_{\mathrm{NLO}_1}~(\mathcal{O}(\alpha_s^2\alpha^2)), \quad \Sigma_{\mathrm{NLO}_2}~(\mathcal{O}(\alpha_s\alpha^3)),\quad \Sigma_{\mathrm{NLO}_3}~(\mathcal{O}(\alpha^4))\,.
\end{equation}

Since we aim to compute all of the above contributions,
we must, for the same reasons explained in ref.~\cite{Frederix:2018nkq}, include massless charged leptons, electrons ({\tt e-}), muons ({\tt mu-}), and taus ({\tt ta-}) as well as their antiparticles and photons ({\tt a}) in the jet ({\tt j}) definition. We treat all these massless particles (leptons, quarks, gluons, and photons) on an equal footing (\ie\ democratic jets) and cluster them using the anti-$k_T$ algorithm with a jet radius of $R=0.4$, applying the cuts in eq.~\eqref{eq:jetcuts}, as done similarly in inclusive dijet hadroproduction in ref.~\cite{Frederix:2016ost}.

\begin{figure}[h!]
% \begin{center}
  \begin{tikzpicture}[line width=1 pt, scale=0.9]
  \hspace{0cm}
\draw[photon] (-2,2.0) -- (1,2.0);
\draw[photon] (-2,-2.0) -- (1,-2.0);
\draw[fermion] (4,2.0) -- (1,2.0);
\draw[fermion] (1,2.0) -- (1,0.7);
\draw[fermion] (1,0.7) -- (4,0.7);
\draw[vector, segment length=6pt] (1,-0.7) -- (1,0.7);
%\draw[gluon] (1,-0.7) -- (1,0.7);
\draw[top]  (4,-2.0) -- (1,-2.0);
\draw[top] (1,-2.0) -- (1,-0.7);
\draw[top] (1,-0.7) -- (4,-0.7);
\node at (-2.3,2.0) {\Large $\gamma$};
\node at (-2.3,-2.0) {\Large $\gamma$};
\node at (4.3,-2.0) {\Large $\bar{t}$};
\node at (4.3,-0.7) {\Large $t$};
\node at (4.3,2.0) {\Large $\bar{f}$};
\node at (4.3,0.7) {\Large $f$};
\node at (2.1,0) {\Large $g/\gamma/Z$};
\hspace{8cm}
\draw[photon] (-2,2.0) -- (0,2.0);
\draw[photon] (-2,-2.0) -- (0,-2.0);
\draw[fermion] (4,2.0) -- (0,2.0);
\draw[fermion] (0,2.0) -- (0,-2.0);
\draw[fermion] (0,-2.0) -- (4,-2.0);
\draw[vector, segment length=6pt] (0,0.0) -- (1.8,0.0);
\draw[top]  (4,-1.0) -- (1.8,0.0);
\draw[top] (1.8,0.0) -- (4,1.0);
\node at (-2.3,2.0) {\Large $\gamma$};
\node at (-2.3,-2.0) {\Large $\gamma$};
\node at (4.3,-1.0) {\Large $\bar{t}$};
\node at (4.3,1.0) {\Large $t$};
\node at (4.3,2.0) {\Large $\bar{f}$};
\node at (4.3,-2.0) {\Large $f$};
\node at (0.9,0.5) {\Large $g/\gamma/Z$};
\end{tikzpicture}
\caption{Representative Feynman diagrams of real-emission contributions to the process $\gaga\to \ttbar j$ at NLO, which exhibit initial-state collinear singularities. The internal wavy line can represent a gluon, a photon, or a $Z$ boson.} 
\label{fig:ttbarjfeynmandiag}
\end{figure}
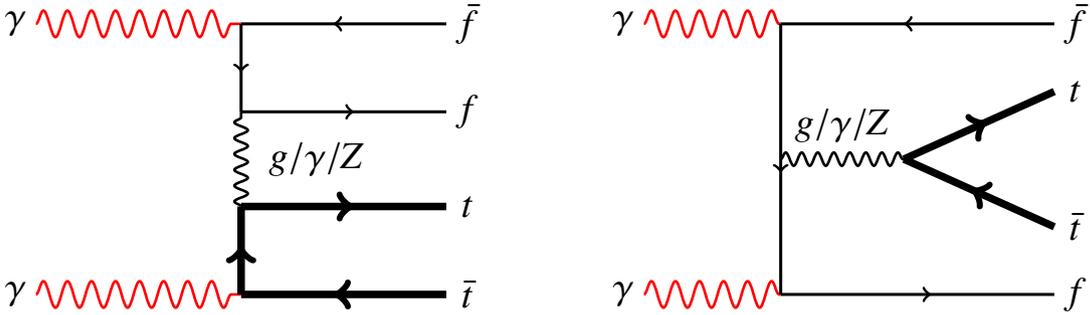

To ensure initial-state collinear safety (cf. section \ref{sec:theory}), the parton content of the initial-state coherent photons must include all massless charged fermions and photons, which are embedded in the jet definition. In addition, the PDF counterterm in eq.~\eqref{eq:Kdef4photon} must be introduced by setting {\tt pdfscheme = 7} in the {\tt run\_card.dat}. Consequently, we vary the parameter $\xi_{\mathrm{A}}$ as an additional source of theoretical uncertainty, taking a central value of unity and varying it up and down by a factor of two. Two representative real-emission Feynman diagrams with initial-state collinear singularities from $\gamma\to f\bar{f}$ splitting are shown in figure \ref{fig:ttbarjfeynmandiag}.

\begin{table}[htpb!]
%\centering
\tabcolsep=3.5mm
%\caption{CAPTION}
%\label{tab:LABEL}
\vspace{0.2cm}
\hspace{-0.8cm}
\begin{tabular}{l|c|c|c} \hline
Process: $\gaga\to \ttbar j$ & \multicolumn{3}{c}{\gammaUPC+\mgshort} \\
Colliding system, \cm\ energy & $\sigma_{\mathrm{LO}_1}$ & $\sigma_{\mathrm{NLO}\ \mathrm{QCD}}$ & $\sigma_{\mathrm{NLO}}$ \\ \hline
\pp\ at 13~TeV & $18.89(1)^{+1.83}_{-1.53}$ ab & $25.78(1)^{+1.31+0.80}_{-1.25-0.73}$ ab & $25.59(1)^{+1.21+0.82}_{-1.17-0.78}$ ab \\
\pp\ at 13.6~TeV & $20.68(1)^{+2.01}_{-1.67}$ ab & $28.36(1)^{+1.47+0.83}_{-1.39-0.85}$ ab & $28.16(1)^{+1.36+0.85}_{-1.30-0.88}$ ab \\
\pp\ at 14~TeV & $21.89(1)^{+2.12}_{-1.77}$ ab & $30.11(2)^{+1.57+0.87}_{-1.49-0.86}$ ab & $29.89(2)^{+1.45+0.90}_{-1.39-0.88}$ ab \\
\pPb\ at 8.8~TeV &  $1.936(1)^{+0.192}_{-0.159}$ fb & $2.479(1)^{+0.103+0.041}_{-0.105-0.035}$ fb & $2.465(1)^{+0.095+0.042}_{-0.098-0.036}$ fb \\
\PbPb\ at 5.52~TeV &  $1.656(1)^{+0.163}_{-0.136}$ fb & $2.225(2)^{+0.109+0.041}_{-0.106-0.041}$ fb & $2.215(2)^{+0.102+0.042}_{-0.099-0.043}$ fb \\
\pp\ at 100~TeV & $334.5(1)^{+31.8}_{-26.6}$ ab & $545.6(9)^{+40.8+22.4}_{-35.5-25.0}$ ab & $542.1(9)^{+38.6+23.4}_{-33.7-25.9}$ ab \\
\pPb\ at 62.8~TeV & $320.5(1)^{+30.9}_{-25.8}$ fb & $472.3(4)^{+29.3+13.3}_{-26.6-13.2}$ fb & $469.6(4)^{+27.5+13.8}_{-25.1-13.8}$ fb \\
\PbPb\ at 39.4~TeV & $46.87(2)^{+4.63}_{-3.85}$ pb & $62.39(6)^{+2.98+1.06}_{-2.90-1.18}$ pb & $62.09(6)^{+2.77+1.09}_{-2.73-1.19}$ pb \\
\hline
 $K$ factor & & $\sigma_{\mathrm{NLO}\ \mathrm{QCD}}/\sigma_{\mathrm{LO}_1}$ & $\sigma_{\mathrm{NLO}}/\sigma_{\mathrm{LO}_1}$ \\ \hline
\pp\ at 13~TeV & & 1.364 & 1.355 \\
\pp\ at 13.6~TeV & & 1.371 & 1.362 \\
\pp\ at 14~TeV & & 1.375 & 1.365 \\
\pPb\ at 8.8~TeV & & 1.281 & 1.273 \\
\PbPb\ at 5.52~TeV & & 1.344 & 1.338 \\
\pp\ at 100~TeV & & 1.631 & 1.620 \\
\pPb\ at 62.8~TeV & & 1.474 & 1.465 \\
\PbPb\ at 39.4~TeV & & 1.331 & 1.325 \\ \hline
\end{tabular}
\caption{The integrated fiducial cross sections (within the fiducial cuts defined in section \ref{sec:ttbarjet}) for $\gaga\to \ttbar j$ at LO$_1$, NLO QCD, and complete NLO accuracy at LHC and FCC energies, computed using the ChFF $\gamma$ fluxes available in \gammaUPC. The quoted uncertainties in the absolute cross sections arise
from variations of the renormalization scale and $\xi_{\mathrm{A}}$ by a factor of $2$, where the latter applies only to the NLO cross sections. Numbers in parentheses indicate statistical errors from numerical integrations and may affect the last decimal place.\label{tab:ttbarjxs}}
\end{table}

\begin{table}[htpb!]
\centering
\tabcolsep=3.5mm
%\caption{CAPTION}
%\label{tab:LABEL}
\vspace{0.2cm}
\begin{tabular}{l|c|c|c} \hline
Process: $\gaga\to \ttbar j$ & \multicolumn{3}{c}{\gammaUPC+\mgshort} \\
Colliding system, \cm\ energy & $\Sigma_{\mathrm{LO}_1}$ & $\Sigma_{\mathrm{LO}_2}$ & \\ \hline
\pp\ at 13~TeV & $18.89(1)$ ab & $0.476(1)$ ab & \\
\pp\ at 13.6~TeV & $20.68(1)$ ab & $0.521(1)$ ab & \\
\pp\ at 14~TeV & $21.89(1)$ ab & $0.552(1)$ ab & \\
\pPb\ at 8.8~TeV &  $1.936(1)$ fb & $0.0478(1)$ fb & \\
\PbPb\ at 5.52~TeV &  $1.656(1)$ fb & $0.0410(2)$ fb & \\
\pp\ at 100~TeV & $334.5(1)$ ab & $8.56(4)$ ab & \\
\pPb\ at 62.8~TeV & $320.5(1)$ fb & $8.13(1)$ fb & \\
\PbPb\ at 39.4~TeV & $46.87(2)$ pb & $1.17(1)$ pb \\
\hline
Colliding system, \cm\ energy & $\Sigma_{\mathrm{NLO}_1}$ & $\Sigma_{\mathrm{NLO}_2}$ & $\Sigma_{\mathrm{NLO}_3}$ \\ \hline
\pp\ at 13~TeV & $6.88(1)$ ab & $-1.05(1)$ ab & $0.385(2)$ ab \\
\pp\ at 13.6~TeV & $7.68(1)$ ab & $-1.15(1)$ ab & $0.434(2)$ ab \\
\pp\ at 14~TeV & $8.21(2)$ ab & $-1.23(1)$ ab & $0.459(1)$ ab\\
\pPb\ at 8.8~TeV &  $0.543(1)$ fb & $-0.0846(1)$ fb & $0.0224(1)$ fb \\
\PbPb\ at 5.52~TeV &  $0.570(2)$ fb & $-0.0781(3)$ fb & $0.0264(2)$ fb \\
\pp\ at 100~TeV & $211(1)$ ab & $-24.2(1)$ ab & $12.1(1)$ ab \\
\pPb\ at 62.8~TeV & $152(1)$ fb & $-19.3(1)$ fb &  $8.50(4)$ fb\\
\PbPb\ at 39.4~TeV & $15.5(1)$ pb & $-2.15(1)$ pb & $0.677(4)$ pb \\
\hline
\end{tabular}
\caption{Breakdown of the integrated fiducial cross sections (within the fiducial cuts defined in section \ref{sec:ttbarjet}) for $\gaga\to \ttbar j$ into five coupling orders at LHC and FCC energies.\label{tab:ttbarjxsbreakdown}}
\end{table}

Table \ref{tab:ttbarjxs} presents the phase-space integrated fiducial cross sections for the process $\gaga\to \ttbar j$ at four, two, and two \cm\ energies in \pp, \pPb, and \PbPb\ collisions at the LHC and FCC-hh. As anticipated, these cross sections are approximately one order of magnitude smaller than those of $\gaga\to \ttbar$ (cf. table \ref{tab:ttbarxs}), due to the $\mathcal{O}(\alpha_s)$ suppression. The NLO QCD corrections significantly enhance the LO cross sections, ranging from a $28\%$ increase at $\sqrtsnn=8.8$ TeV in \pPb\ collisions at the LHC to a $63\%$ increase at $\sqrts=100$ TeV in \pp\ collisions at the FCC-hh. These enhancements are generally larger than those observed in $\gaga\to \ttbar$. The complete NLO cross sections, which include all five coupling orders, are found to be close to the NLO QCD cross sections. This proximity is not due to  the insignificance of the LO$_2$, NLO$_2$, and NLO$_3$ terms, but rather due to their mutual cancellations, as clearly shown in table \ref{tab:ttbarjxsbreakdown}. In particular, $\Sigma_{\mathrm{NLO}_2}$ (the conventional NLO EW correction) reduces $\Sigma_{\mathrm{LO}_1}$ by $4$-$7\%$, while $\Sigma_{\mathrm{LO}_2}$ increases it by around $2.5\%$, and $\Sigma_{\mathrm{NLO}_3}$ enhances it by $2$-$4\%$, depending on the colliding system and \cm\ energy. The combined effect of these subleading contributions, $\Sigma_{\mathrm{LO}_2}+\Sigma_{\mathrm{NLO}_2}+\Sigma_{\mathrm{NLO}_3}$, modifies the integrated cross section by approximately $-1\%$ overall. Furthermore, $\Sigma_{\mathrm{NLO}_2}$ is approximately two to four times larger than $\Sigma_{\mathrm{NLO}_3}$ in absolute terms, which is consistent with simple power-counting arguments based on the hierarchy $\alpha\ll \alpha_s$.

Together with the central values, the theoretical uncertainties are also presented in table \ref{tab:ttbarjxs}. These uncertainties arise from two distinct sources: the first from the variation of the renormalization scale $\mu_R$, and the second from the variation of $\xi_{\mathrm{A}}$, both by a factor of 2. Naturally, the uncertainty associated with $\xi_{\mathrm{A}}$ appears only in the NLO cross sections. The fractional renormalization scale dependence is reduced from approximately $\pm 10\%$ at LO to $\pm 5\%$ at LHC energies and $\pm 7\%$ at $\sqrt{s}=100$ TeV. Clearly, the LO scale uncertainty significantly underestimates the actual size of the NLO QCD corrections. In comparison, the uncertainty due to $\xi_{\mathrm{A}}$ variation is subdominant. The largest fractional $\xi_{\mathrm{A}}$ uncertainty is around $\pm 4\%$ at 100 TeV in the \pp\ collision mode, while for other collision systems and \cm\ energies, it ranges between $\pm 2\%$ and $\pm 3\%$. It remains to be seen whether the combined scale and $\xi_{\mathrm{A}}$ uncertainties at NLO capture the bulk of the missing higher-order corrections.

\subsection{$\gaga\to \fourmuon$\label{sec:fourmuon}}

The four-muon production process $\gaga\to \fourmuon$ constitutes an irreducible background for the loop-induced process $\gaga\to ZZ$, where each $Z$ boson decays into a muon pair, as well as for the exclusive production of a pair of $J/\psi$ mesons. The $Z$-boson pair production process is very rare (see the LO cross sections in table IX of ref.~\cite{Shao:2022cly}) since it is a loop-induced process in the SM, but it is interesting for probing the anomalous quartic gauge couplings. The central-exclusive production of double $J/\psi$ mesons has been observed by the LHCb collaboration~\cite{LHCb:2014zwa} in \pp\ at $\sqrt{s}=7$ and 8 TeV. In \pp\ collisions, the gluon-induced central exclusive production is dominant over the photon-induced process $\gaga\to J/\psi J/\psi$. 

We generate $\gaga\to\fourmuon$ at NLO as follows:
\vskip 0.25truecm
\noindent
~~\prompt\ {\tt ~set~complex\_mass\_scheme~True}

\noindent
~~\prompt\ {\tt ~import~model~loop\_qcd\_qed\_sm\_Gmu\_3FS-a0}

\noindent
~~\prompt\ {\tt ~generate \taggeda\ \taggeda\ > mu+ mu+ mu- mu- [QED]}

\vskip 0.25truecm
\noindent
We use the complex-mass scheme due to the presence of the OS $Z$ boson, where only one OS $Z$ can exist. The five-flavor number scheme is adopted here. All leptons are considered massless, with the muon being dressed. The process features $k_0=4$, $c_s(k_0)=0$, $c(k_0)=4$, and $\Delta(k_0)=0$, as defined in eq.~\eqref{eq:Sigmaexp0}, with the NLO QCD correction term $\Sigma_{\mathrm{NLO}_1}$ being zero.

\begin{table}[t]
\centering
\tabcolsep=3.5mm
\vspace{0.2cm}
\begin{tabular}{l|c|c|c} \hline
Process: $\gaga\to \fourmuon$ & \multicolumn{3}{c}{\gammaUPC+\mgshort} \\
Colliding system, \cm\ energy & $\sigma_{\mathrm{LO}}$ & $\sigma_{\mathrm{NLO}\ \mathrm{EW}}$ & $\sigma_{\mathrm{NLO}\ \mathrm{EW}}/\sigma_{\mathrm{LO}}$ \\ \hline
\pp\ at 13~TeV &  $1.604(2)$ ab &  $1.610(2)$ ab & 1.004 \\
\pp\ at 13.6~TeV &  $1.674(2)$ ab &  $1.679(2)$ ab & 1.003 \\
\pp\ at 14~TeV &  $1.718(4)$ ab &  $1.723(4)$ ab & 1.003 \\
%\pPb\ at 8.8~TeV &  $20.73(2)$ ab &  $20.24(2)$ ab & 0.976 \\
\pPb\ at 8.8~TeV &  $0.7924(6)$ fb &  $0.7860(8)$ fb & 0.992 \\
\PbPb\ at 5.52~TeV &  $0.1541(2)$ pb &  $0.1495(2)$ pb & 0.970 \\
\hline
\pp\ at 100~TeV &  $6.377(7)$ ab &  $6.423(8)$ ab & 1.007 \\
\pPb\ at 62.8~TeV &  $12.45(1)$ fb &  $12.50(2)$ fb & 1.004 \\
\PbPb\ at 39.4~TeV &  $21.69(2)$ pb &  $21.65(2)$ pb & 0.998 \\
\hline
\end{tabular}
\caption{The integrated fiducial cross sections (within the fiducial cuts defined in section \ref{sec:setup}) for $\gaga\to \fourmuon$ at LO and NLO accuracy at LHC and FCC-hh energies, computed using the ChFF $\gamma$ fluxes available in \gammaUPC. The last column also shows the corresponding $K$ factors. Numbers in parentheses indicate statistical errors from numerical integrations, which may affect the last decimal place.\label{tab:fourmuonxs}}
\end{table}

The corresponding LO and NLO cross sections, both at LHC and FCC-hh energies, within the fiducial volume defined in section \ref{sec:setup}, can be found in table \ref{tab:fourmuonxs}. This is a very rare process, with the \pp\ cross sections below $2$ ab ($7$ ab) at the LHC (FCC-hh). The cross sections in \pPb\ are roughly three orders of magnitude larger than in \pp, due to the $Z^2$ enhancement of the photon-photon luminosities. Similarly, the cross section in \PbPb\ is approximately 200 (1,700) times larger than in \pPb\ at the LHC (FCC-hh). The NLO EW corrections are negative and largest in absolute terms in \PbPb\ collisions at $\sqrtsnn=5.52$ TeV, while those at other \cm\ energies or in other colliding systems are at sub-percent level and can be either positive or negative.

\subsection{$\gaga\to \llvv$\label{sec:llvv}}

We study the off-shell effect for the photon-fusion process $\gaga\to \WpWm \to \llvv$ at NLO using the complex-mass scheme, which can be generated through:
\vskip 0.25truecm
\noindent
~~\prompt\ {\tt ~set~complex\_mass\_scheme~True}

\noindent
~~\prompt\ {\tt ~import~model~loop\_qcd\_qed\_sm\_Gmu\_3FS-a0}

\noindent
~~\prompt\ {\tt ~generate \taggeda\ \taggeda\ > mu+ vm e- ve\~{} [QED]}

\vskip 0.25truecm
\noindent
Similar to $\gaga\to\fourmuon$ presented in section \ref{sec:fourmuon}, the process has $k_0=4$, $c_s(k_0)=0$, $c(k_0)=4$, and $\Delta(k_0)=0$, with vanishing $\Sigma_{\mathrm{NLO}_{1}}$. However, unlike $\gaga\to\fourmuon$, the cross section for $\gaga\to \llvv$ is dominated by the phase space regime of two OS $W$ bosons.

\begin{table}[t]
\centering
\tabcolsep=3.5mm
\vspace{0.2cm}
\begin{tabular}{l|c|c|c} \hline
Process: $\gaga\to \llvv$ & \multicolumn{3}{c}{\gammaUPC+\mgshort} \\
Colliding system, \cm\ energy & $\sigma_{\mathrm{LO}}$ & $\sigma_{\mathrm{NLO}\ \mathrm{EW}}$ & $\sigma_{\mathrm{NLO}\ \mathrm{EW}}/\sigma_{\mathrm{LO}}$ \\ \hline
\pp\ at 13~TeV &  $0.4604(1)$ fb &  $0.4548(2)$ fb & 0.988 \\
\pp\ at 13.6~TeV &  $0.4850(2)$ fb &  $0.4787(3)$ fb & 0.987 \\
\pp\ at 14~TeV &  $0.5012(1)$ fb &  $0.4943(2)$ fb & 0.986 \\
\pPb\ at 8.8~TeV &  $0.1750(1)$ pb &  $0.1755(1)$ fb & 1.003 \\
\PbPb\ at 5.52~TeV &  $3.496(1)$ pb &  $3.556(2)$ pb & 1.017 \\
\hline
\pp\ at 100~TeV &  $2.393(1)$ fb &  $2.323(1)$ fb & 0.971 \\
\pPb\ at 62.8~TeV &  $3.839(1)$ pb &  $3.787(3)$ pb & 0.987 \\
\PbPb\ at 39.4~TeV &  $4.390(2)$ nb &  $4.382(2)$ nb & 0.998 \\
\hline
\end{tabular}
\caption{The integrated fiducial cross sections (within the fiducial cuts defined in section \ref{sec:setup}) for $\gaga\to \llvv$ at LO and NLO accuracy at LHC and FCC-hh energies, computed using the ChFF $\gamma$ fluxes available in \gammaUPC. The last column also shows the corresponding $K$ factors. Numbers in parentheses indicate statistical errors from numerical integrations, which may affect the last decimal place.\label{tab:llvvxs}}
\end{table}

The phase-space integrated fiducial cross sections are summarized in table \ref{tab:llvvxs}. The absolute cross sections are similar to the inclusive cross sections for producing two OS $W$ bosons, multiplied by their decay branching ratio $\mathrm{Br}(\WpWm\to \llvv)\approx 1.14\%$~\cite{ParticleDataGroup:2024cfk} and the phase-space acceptance factor due to the imposed fiducial cuts. The NLO EW corrections are generally negative, except for the cases of \pPb\ at $\sqrtsnn=8.8$ TeV and \PbPb\ at $\sqrtsnn=5.52$ TeV. The overall corrections modify the LO cross sections at LHC and FCC-hh energies from $-3\%$ to $+2\%$. However, these correction sizes could be enhanced in the high-energy tails of differential distributions due to the presence of EW Sudakov logarithms. We refrain from showing any differential cross sections here, but interested readers can generate them using our public program, which will be released in the near future.

\section{Conclusion\label{sec:conclusion}}

In this paper, we have extended the \mgshort\ framework to enable automated computations of NLO QCD and/or EW corrections for arbitrary final states $X$ in photon-photon collisions within UPCs of protons and ions, generically denoted as \ABgagaX. The effective photon-photon luminosities are computed using the \gammaUPC\ code, which has been integrated into \mgshort. Several technical aspects of the implementation are discussed, including a mixed EW renormalization scheme and modifications to the FKS subtraction method for processes with or without jets. A summary of the available computations within this framework is provided in table \ref{tab:allowedinMG}.

Some illustrative examples of photon-photon cross sections computed automatically within the \gammaUPC+\mgshort\ framework have been presented for proton-proton, proton-nucleus, and nucleus-nucleus UPCs at the LHC and FCC-hh. We provide results for total inclusive, integrated fiducial, and/or differential cross sections of dilepton production (for both dressed and bare leptons), $\WpWm$, top-antitop quark pairs, the associated processes involving $\ttbar$ with an isolated tagged photon or a democratic jet, as well as four-muon and two-lepton-two-neutrino final states. These processes not only demonstrate the computational capabilities of the framework but are also of significant phenomenological interest. They serve as standard candles for calibrating (coherent) photon fluxes and offer valuable opportunities for novel SM tests--such as measurements of $\tau$ lepton and top-quark electromagnetic moments and quartic gauge couplings--while also opening new avenues for BSM searches.

\begin{acknowledgments}
We are grateful to David d'Enterria for reading the manuscript and providing valuable feedback. This work is supported by the ERC (grant 101041109 ``BOSON") and the French ANR (grant
ANR-20-CE31-0015, ``PrecisOnium"). Views and opinions expressed are however those of the authors only and do not necessarily reflect those of the European Union or the European Research Council Executive Agency. Neither the European Union nor the granting authority can be held responsible for them.
\end{acknowledgments}

\bibliographystyle{myutphys}
\bibliography{reference}

\end{document}